\documentclass[sn-mathphys-num,Numbered,twocolumn]{sn-jnl}
\usepackage{threeparttable}
\usepackage{graphicx}%
\usepackage{multirow}%
\usepackage{amsmath,amssymb,amsfonts}%
\usepackage{amsthm}%
\usepackage{mathrsfs}%
\usepackage[title]{appendix}%
\usepackage{xcolor}%
\usepackage{textcomp}%
\usepackage{manyfoot}%
\usepackage{booktabs}%

\usepackage{algorithm}%
\usepackage{algpseudocode}%

\usepackage{listings}%
\usepackage{ifpdf}
\usepackage{caption}
\usepackage{subcaption}
\usepackage{booktabs}
\usepackage{graphicx}
\usepackage{hyperref}
\usepackage{placeins}
\usepackage{multirow}
\usepackage{float}
\usepackage[T1]{fontenc}
\usepackage[utf8]{inputenc}
\usepackage{xcolor}
\usepackage{url}
\usepackage{hyphenat}
\usepackage{adjustbox}
\usepackage{array}
\usepackage{color}
\usepackage{url}
\usepackage[normalem]{ulem}
\usepackage{footnote}
\usepackage{lipsum}
\usepackage{multirow}
\usepackage{hyperref}
\usepackage{comment}
\usepackage{placeins}
\usepackage{afterpage}
\usepackage{appendix}
\usepackage{lscape} 
\usepackage{amsmath}
\usepackage{amssymb}
\usepackage{float}

\definecolor{dkgreen}{rgb}{0,0.6,0}
\definecolor{gray}{rgb}{0.5,0.5,0.5}
\definecolor{mauve}{rgb}{0.58,0,0.82}

\newcolumntype{R}[2]{%
    >{\adjustbox{angle=#1,lap=\width-(#2)}\bgroup}%
    l%
    <{\egroup}%
}

\begin{document}




\title{DNS-GT: A Graph-based Transformer Approach to Learn Embeddings of Domain Names from DNS Queries}

\author[1]{\fnm{Massimiliano} \sur{Altieri}}\email{massimiliano.altieri@ec.europa.eu}
\author[1]{\fnm{Ronan} \sur{Hamon}}\email{ronan.hamon@ec.europa.eu}
\author*[2]{\fnm{Roberto} \sur{Corizzo}}\email{rcorizzo@american.edu}
\author[3]{\fnm{Michelangelo} \sur{Ceci}}\email{michelangelo.ceci@uniba.it}
\author*[1]{\fnm{Ignacio} \sur{Sanchez}}\email{ignacio.sanchez@ec.europa.eu}

\affil[1]{\orgdiv{Joint Research Centre}, \orgname{European Commission}, \orgaddress{ \city{Ispra}, \postcode{21027}, \country{Italy}}}

\affil[2]{\orgdiv{Department of Computer Science}, \orgname{American University}, \orgaddress{\street{4400 Massachusetts Avenue NW}, \city{Washington}, \postcode{20016}, \state{DC}, \country{USA}}}

\affil[3]{\orgdiv{Department of Computer Science}, \orgname{University of Bari Aldo Moro}, \orgaddress{\street{Via Orabona, 4}, \city{Bari}, \postcode{70125}, \country{Italy}}}


\abstract{
Network intrusion detection systems play a crucial role in the security strategy employed by organisations to detect and prevent cyberattacks. Such systems usually combine pattern detection signatures with anomaly detection techniques powered by machine learning methods. However, the commonly proposed machine learning methods present drawbacks such as over-reliance on labeled data and limited generalization capabilities. To address these issues, embedding-based methods have been introduced to learn representations from network data, such as DNS traffic, mainly due to its large availability, that generalise effectively to many downstream tasks. However, current approaches do not properly consider contextual information among DNS queries. In this paper, we tackle this issue by proposing DNS-GT, a novel Transformer-based model that learns embeddings for domain names from sequences of DNS queries. The model is first pre-trained in a self-supervised fashion in order to learn the general behavior of DNS activity. Then, it can be finetuned on specific downstream tasks, exploiting interactions with other relevant queries in a given sequence. 
Our experiments with real-world DNS data showcase the ability of our method to learn effective domain name representations. A quantitative evaluation on domain name classification and botnet detection tasks shows that our approach achieves better results compared to relevant baselines, creating opportunities for further exploration of large-scale language models for intrusion detection systems. 
Our code is available at: \url{https://github.com/m-altieri/DNS-GT}.
}


\maketitle


\clearpage

\section{Introduction}
\label{sec:introduction}

{

Network intrusion detection systems (NIDS) are essential components of modern cybersecurity infrastructures. They monitor network traffic to detect malicious activities such as botnet communications, remote exploitation of services, or Denial of Service (DoS) attacks. These systems typically combine signature-based methods to detect known threats and anomaly detection techniques to detect novel attacks.

Cyberattacks inevitably leave traces in network activity, whether through abnormal host behaviors, data exfiltration patterns, or signs of service disruption~\cite{ENISA2022Threat}. As attacks grow more sophisticated, traditional rule-based systems appear inadequate. In response, machine learning (ML) techniques have gained traction for intrusion detection~\cite{ahmad2021network}, offering the ability to identify complex patterns in large volumes of traffic and adapt to evolving threats~\cite{khraisat2019survey}.

This shift toward ML-based detection has been further fueled by the success of large-scale models in fields such as computer vision and natural language processing (NLP)~\cite{Aghaei2023SecureBERT, Akbar2022Knowledge}. However, a major bottleneck in applying these models to cybersecurity is the limited availability of large-scale datasets. Privacy concerns surrounding user data exacerbate this limitation, restricting the availability of labelled  data to train general models~\cite{Buczak2016survey}.

To address these challenges, some approaches focus on DNS (Domain Name System) traffic, a protocol that maps domain names to IP addresses, as a data source for intrusion detection. 
%
DNS traffic offers a partial yet comprehensive view of network activity, covering diverse applications. It is relatively simple, making it often available in large volumes within organizations. DNS-based monitoring can also detect threats early, before malicious requests reach their targets.

Recent works leveraged DNS data to learn domain name embeddings using deep learning methods adapted from NLP, such as Word2Vec~\cite{lopez2017vector, lopez2020learning, Morbidoni2022Leveraging}. These embeddings, akin to word vectors in NLP, can serve as representations for various cybersecurity tasks~\cite{roy2017learning}. However, existing models based on Word2Vec variants lack a deep understanding of context and semantics, since they aggregate local co-occurrence patterns. 

Transformer-based models, widely successful in NLP, address this limitation by leveraging self-attention mechanisms that capture complex dependencies in input sequences. Inspired by this, we propose DNS-GT, a novel Transformer architecture incorporating graph neural modeling tailored for cybersecurity.

\clearpage\newpage
DNS-GT learns contextual representations of domain names and network hosts from raw, unlabeled DNS data using a masked language modeling objective during pre-training. It can then be fine-tuned for specific downstream tasks, even with limited labeled data.
Unlike earlier embedding methods, DNS-GT's multi-head attention and graph neural layers allow it to selectively focus on task-relevant contextual information, enhancing its performance across varied cybersecurity applications.

Our main contributions are as follows:
\begin{itemize}
    \item We introduce DNS-GT, a novel Transformer model with integrated graph neural modeling that learns robust embeddings from DNS data;
    \item We conduct extensive experiments on a real-world DNS traffic dataset with over 4,000 hosts, providing both qualitative and quantitative evaluations;
    \item We demonstrate the model's generalizability by applying it to domain classification and botnet detection as downstream tasks.
\end{itemize}

The rest of the paper is organized as follows: Section~\ref{sect:related-work} reviews related work on intrusion detection and DNS-based methods. Section~\ref{sec:method} details our methodology and model architecture. Section~\ref{sec:experiments} presents the experimental setup and evaluation, and Section~\ref{sec:conclusion} concludes the paper.
}

\section{Related Work}
\label{sect:related-work}

{
Network intrusion detection systems typically rely on analysing features from network communications across different OSI layers~\cite{CyberEdu2018What}, including packet-level attributes (e.g., IPs, ports, flags, payloads) and session-level characteristics (e.g., duration, packet counts). Traditional methods use signature-based detection via handcrafted rules by security analysts. However, with the increasing attack volume and complexity, ML-based anomaly detection has gained traction~\cite{ahmad2021network,10190520}, offering automatic adaptation to new threats by learning complex patterns from large data~\cite{khraisat2019survey}.


Early ML methods like SVMs or random forests~\cite{hastie2009support}, and fuzzy logic\cite{kim2024flsec} methods, used handcrafted features, which, while effective~\cite{di2021supervised, fares2011intrusion, wang2004anomalous}, have been largely replaced by neural networks that are able to learn features from raw data. However, this shift improves pattern detection but reduces interpretability~\cite{xie2013evaluating} 
\cite{xu2024procsage}. Additionally, both approaches require labeled data, which is costly and often incomplete, and depend heavily on the available features, which may become outdated. Semi-supervised methods, especially autoencoders and one-class models~\cite{choi2019unsupervised, binbusayyis2021unsupervised, vaiyapuri2020application,faber2022active}, can help mitigate these issues by detecting distribution shifts and by being feature-agnostic.


DNS traffic is widely used in malicious domain detection due to its high availability and scope~\cite{zhauniarovich2018survey}. In the context of DNS data, earlier methods rely on handcrafted features for tasks like anomaly-based botnet detection~\cite{choi2007botnet, grill2015detecting}, but require constant updates, limiting scalability. ML approaches, in contrast, can uncover hidden dependencies without predefined rules. Supervised techniques have been applied to labeled DNS traffic to detect malicious patterns~\cite{antonakakis2011detecting}, though labeling remains a bottleneck. To address this, unsupervised methods using domain name embeddings have emerged, leveraging character-level models~\cite{Morbidoni2022Leveraging}, Word2Vec~\cite{lopez2017vector}, and other embedding methods like fastText and app2vec~\cite{lopez2020learning}.

More recently, hybrid approaches that integrate temporal dynamics with structural information have gained attention in cybersecurity applications. This trend includes the use of spatio-temporal graphs and attention-based architectures~\cite{wang2020defending,wu2022graphbert}, which demonstrate how embedding both the temporal progression and contextual structure of data can enhance detection capabilities across diverse threat landscapes.

Building on these insights, our work relies on an attention-based architecture~\cite{Vaswani2017Attention} and graph neural networks (GNNs)~\cite{DBLP:journals/corr/KipfW16} to model the semantics of sequences of concurrently resolved DNS queries, representing either benign or malicious behaviours. The attention mechanism provides context-aware representations across queries, while GNNs incorporate domain knowledge via graph structures that link related queries, ensuring information flows only among relevant nodes. The resulting embeddings capture behavioural patterns among domain names, supporting applications like classification, phishing detection, behaviour profiling, and domain impersonation detection, making DNS traffic analysis a key component in future security strategies.
}

\section{Method}
\label{sec:method}

\begin{figure*}
    \centering
    \includegraphics[width=\linewidth,trim=25 10 20 10,clip]{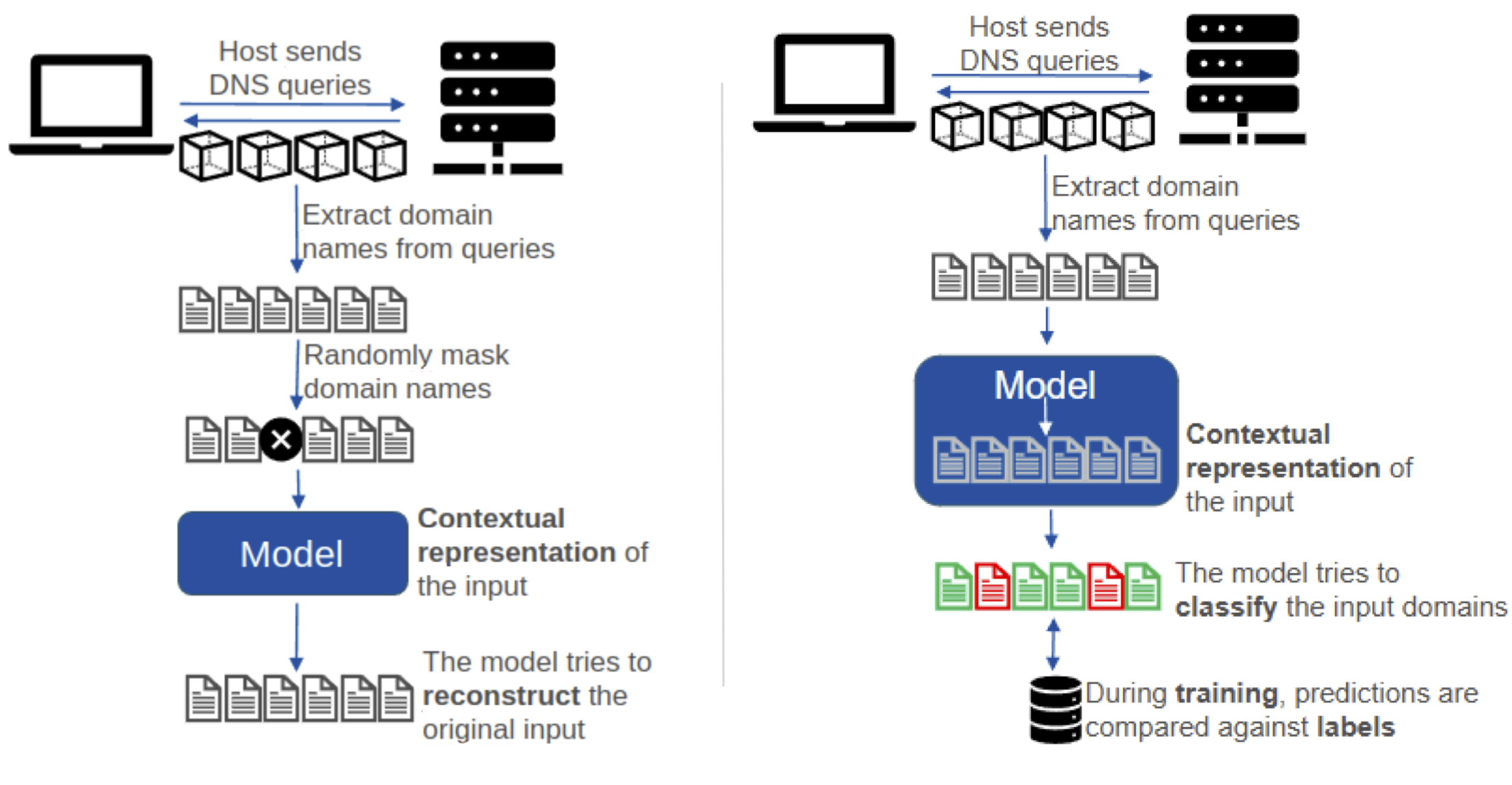} \\
     (a) \ \ \ \ \ \ \ \ \ \ \ \ \ \ \ \ \ \ \ \ \ \ \ \ \ \ \ \ \ (b) \\
    \caption{Conceptual overview of domain classification task with model training phases: pre-training (a) and fine-tuning (b).}
    \label{fig:conceptual-overview-model-phases}
\end{figure*}

\begin{figure*}
    \centering
    \includegraphics[width=0.999\linewidth]{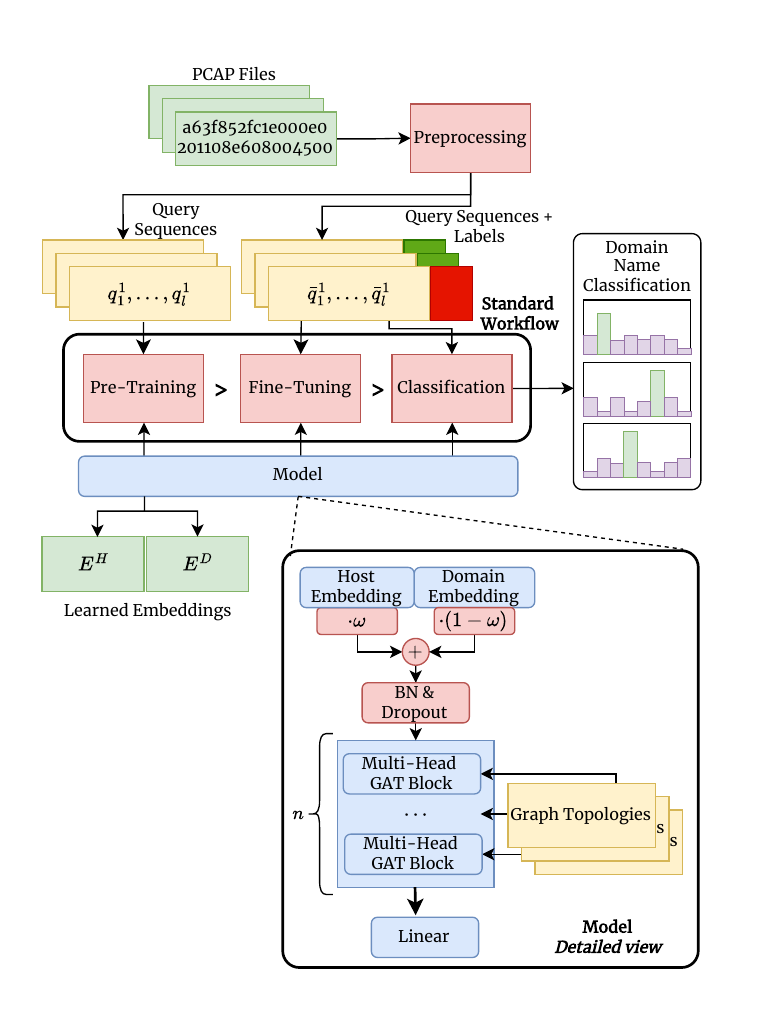}
    \caption{Workflow of the proposed DNS-GT method. Each stage is represented in red. Input and output data are depicted in green.}
    \label{fig:workflow}
\end{figure*}

In this section, we describe our method providing technical details of each stage, including data processing, and we describe details of the proposed model architecture, as well as possible applications \footnote{For a streamlined presentation, we provide preliminary concepts used in our method, such as large language model training stages, attention mechanism, and Graph Attention Networks in our external Appendix: \url{https://github.com/m-altieri/DNS-GT}.}.
Figures~\ref{fig:conceptual-overview-model-phases} and~\ref{fig:workflow} provide high-level visual summaries of the model's pre-training, fine-tuning process, and overall workflow.

Specifically, we propose a novel methodology for network traffic analysis based on DNS queries inspired by recent NLP approaches and, in particular, by the Transformer architecture~\cite{Vaswani2017Attention}. 

The novelty of our approach is twofold. First, it extracts contextual dependencies from neighboring DNS queries, thanks to advanced components that have proven to be effective for several NLP tasks, mainly found in the Transformer \cite{Vaswani2017Attention}, such as self-attention, encoder blocks and residual skip-connections.
Second, several adaptations of the model architecture to the application of network intrusion detection are proposed, such as: \textit{(i)} generalization of the encoder self-attention of the Transformer with a permutation-equivariant graph attention network, \textit{(ii)} dual representation for hosts and domains with a merging mechanism to allow for different use cases, and \textit{(iii)} the possibility to add external knowledge on domain names in the form of graph topologies.

The model is trained using the masked language modeling (MLM) technique~\cite{devlin-etal-2019-bert}, where some input tokens are masked, and the training task consists in correctly identifying the original tokens by assigning a probability score for each token in the vocabulary.

The objective is to update the embeddings and weights of the model to capture the ``grammar'' of DNS query sequences, which is evaluated by its capacity to reconstruct masked domains in the input sequence. 
This work highlights several analogies between NLP and DNS concepts, which hint at the soundness of the approach, along with some differences. 

Our method provides two main advantages:
From the point of view of domain classification and intrusion detection, it provides the opportunity to perform analyses without labeled data. This aspect is particularly important in organizations with a private DNS infrastructure and a large volume of raw network data to be analyzed, where exhaustive data labeling appears unfeasible.

Another important point is transferability with respect to different downstream tasks: the same model pre-trained on raw network data learns contextual representation that can be effectively exploited for domain classification as well as botnet detection.

\subsection{Sequencing}

The raw DNS traffic undergoes a data pre-processing pipeline, aiming at cleaning, filtering, and producing suitable data for training. Due to space constraints, the exact pre-precessing pipeline is described in the Appendix.

The individual DNS queries are packed together into sequences in a cohesive way, which is a crucial step for the correct extraction of DNS patterns by the model, as the attention mechanism is specifically designed to emphasise correlations between tokens belonging to the same sequence.
The timestamp is used for this purpose to ensure that queries from each host are arranged chronologically.

Denoting two queries with \(q_i\) and \(q_j\), \(i<j\), if \(q_i\) chronologically precedes \(q_j\), as a general rule, a sequence \(\mathbf x\) is made of queries that are temporally contiguous:
$$q_i,q_k\in \mathbf x, i\leq j\leq k\implies q_j\in \mathbf x.$$

Beyond this requirement, the specific intra-sequence ordering is ignored by the model, for two reasons: \textit{(i)} the query timestamp is not necessarily a perfect ordering key, due to delays in network transmission, and \textit{(ii)} queries are often performed in ``bursts'', when the host wants to retrieve several resources at once, in which case the overall context of the communication is more important than the exact ordering.

We introduce and evaluate three different strategies for sequencing: 
\begin{enumerate}
    \item \emph{fixed-length}: sequences have a fixed, arbitrary length \(L\) and are obtained by shifting a window by a fixed number of queries \(s\).
    This approach does not take query timing into consideration, but rather only the chronological order, potentially resulting in sequences containing semantically unrelated queries.
    
    \item \emph{greedy time-based}: query timestamp is used to build sequences in a greedy way. For a given sequence with a starting query \(q_0\) with timestamp \(t_0\), a query \(q_i\) with timestamp \(t_i\) is added to the sequence if all these three conditions are met:
    
        (a) the time delay since the previous query timestamp \(t_{i-1}\) is lower than the maximum intra-sequence distance \(\Delta_{\text{intra}}\):
    $$t_i - t_{i-1} < \Delta_{\text{intra}};$$
    
     (b) the base sequence duration \(\Delta\) and the minimum inter-sequence distance \(\Delta_{\text{inter}}\) are not simultaneously exceeded:
    $$t_i - t_0 < \Delta \text{ or } t_i - t_{i-1} < \Delta_{\text{inter}};$$
    
    and (c) the sequence is not full.
    This strategy ensures that a sequence only contains queries that are sent in a rapid succession, and therefore are likely semantically related.
    
    \item \emph{clustering time-based}: creates a partition of the whole query traffic that is performed by a host, using a clustering algorithm along the temporal axis to discover clusters of queries that are temporally close and therefore related. For that purpose, we adopt DBScan~\cite{ester1996density}, a density-based non-parametric clustering algorithm, using the median consecutive time delta as the value for the parameter \(\epsilon\)\footnote{The maximum distance between two samples for one to be considered as in the neighborhood of the other\cite{ester1996density}.}. This approach features robustness to outliers, non-linearity, and a lower computational cost.
\end{enumerate}


\subsection{Knowledge-based Topologies}
\label{sec:topologies}

One of the key aspects of the proposed framework is the possibility to support the graph attention network (GAT) with custom knowledge-based graph topologies, in order to emphasise underlying relationships between tokens in the sequence. 
The graph topologies are fed to the model in the form of adjacency matrices \(A\in\mathbb{R}^{L\times L}\), for sequences of length \(L\), where \(a_{ij}\) represents the connection status between tokens \(i\) and \(j\).
At a high level, such graph topologies are used to constrain the neighborhood of each token to only the subset of related tokens, ignoring the others during the GAT computation.

In our evaluation, we used a simple knowledge-based topology that connects all domains to each other, allowing each (query) token in the attention mechanism to attend to all other (key) tokens.
The only exception is given for the special \texttt{<PAD>} tokens, that are only used as a way to pad sequences to the same length, but should not provide any information to the other domain tokens, and are therefore disconnected from the attention mechanism.

Specifically, given a sequence structured as:
$$\mathbf x = \begin{bmatrix}
    \underbrace{\begin{matrix}q_{1} & q_{2} & \cdots & q_{\ell}\end{matrix}}_{\ell \text{ queries}} &   \underbrace{\begin{matrix}\texttt{<PAD>} & \cdots  & \texttt{<PAD>}\end{matrix}}_{L-\ell\text{ padding}} \\
\end{bmatrix}$$
this behaviour is achieved with the following adjacency matrix:
$$
A =
\begin{bmatrix}
\underbrace{
\begin{matrix}
    1 & 1 & \cdots & 1 \\
    1 & 1 & \cdots & 1 \\
    \vdots & \vdots & \ddots & \vdots \\
    1 & 1 & \cdots & 1 \\
    0 & 0 & \cdots & 0 \\
    0 & 0 & \cdots & 0 \\
    \vdots & \vdots & \ddots & \vdots \\
    0 & 0 & \cdots & 0 \\
\end{matrix}
}_{\ell\text{ queries}}
    &
\underbrace{
\begin{matrix}
     0 & 0 & \cdots & 0 \\
     0 & 0 & \cdots & 0 \\
     \vdots & \vdots & \ddots & \vdots \\
     0 & 0 & \cdots & 0 \\
     1 & 0 & \cdots & 0 \\
     0 & 1 & \cdots & 0 \\
     \vdots & \vdots & \ddots & \vdots \\
     0 & 0 & \cdots & 1 \\
\end{matrix}
}_{L-\ell\text{ padding}}
\end{bmatrix}
$$

However, the proposed model architecture allows to implement and add any kind of domain-specific topologies (e.g., based on the similarity between two domains).

\subsection{Modeling}
\label{sec:modeling}
The model input is a sequence of \(L\) queries \(\mathbf x = [q_1, q_2, \ldots, q_L]\), and each query is a pair of tokens \(q_i=(h_i, d_i)\), where \(h_i\) and \(d_i\) correspond to the querying host token and to the requested domain token, respectively.
Following the MLM approach~\cite{devlin-etal-2019-bert}, each query in the input sequence has at the beginning of the forward pass a fixed probability \(p\) of getting masked.
When a token is masked, it has in turn a probability \(p_{\text{mask}}\) of being replaced with the \texttt{<MASK>} token, a probability \(p_{\text{random}}\) of being replaced with a random token, and a probability \(p_{\text{same}}\) of staying unchanged.


Each host \(h\) and each domain \(d\) is associated with a learnable embedding \(\mathbf e_h^H\) and \(\mathbf e_d^D\), that constitutes a high-dimensional vector representation for that token.
Special tokens like \texttt{<MASK>} also have their respective learned embedding.
Host and domain embeddings are added together to obtain the overall embedding for query \(q_i\), using a fixed weighting 
coefficient \(\omega\in[0,1]\), which determines their degree of contribution:
\(\mathbf e_{q_i}=\omega\cdot \mathbf e_{d_i}^D+(1-\omega)\cdot \mathbf e_{h_i}^H\).
If user privacy is of particular concern, the hyperparameter \(\omega\) allows to easily prevent the model from using any host information, simply by setting \(\omega=1\).
We denote the embedding representation of a sequence with \(\mathbf e = [\mathbf e_{q_1},\ldots,\mathbf e_{q_L}]\). 
After host and domain aggregation, a dropout layer with a rate of \(0.15\) is applied, as well as a batch normalization layer. At a low level, this provides added regularization to the query embeddings, so that the model cannot rely on specific dimensions in the learned embedding to characterize the query, but rather it is forced to utilize each dimension individually, and conversely the learned embeddings will distribute useful information more uniformly among the different dimensions of the vector.

\begin{figure*}
    \centering
    \includegraphics[width=\linewidth]{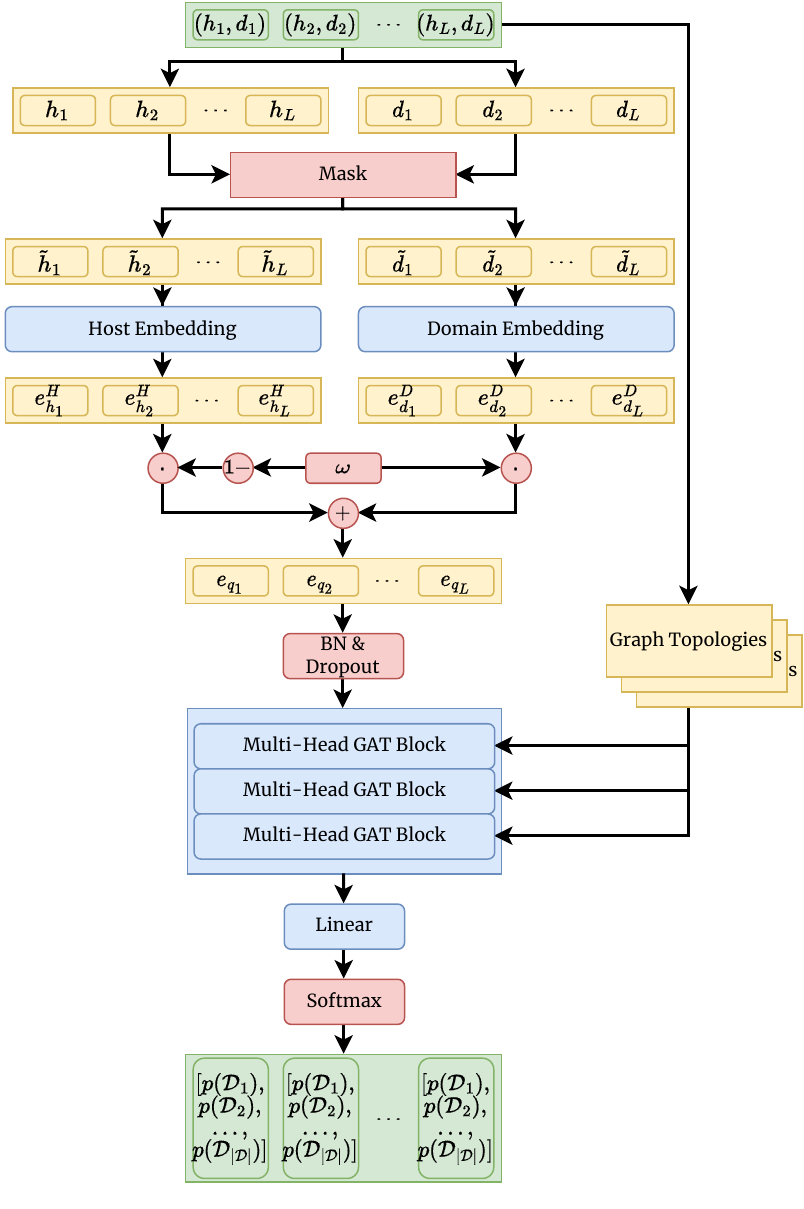}
    \caption{Model architecture. The colour is used to emphasise model input and output (green), tensors (yellow), operations (red) and learnable neural networks (blue).}
    \label{fig:architecture}
\end{figure*}   


The proposed model architecture is based on the Transformer with some key differences, to adapt it to a network cybersecurity context, and a general overview is shown in Figure \ref{fig:architecture}.
The first difference is the absence of a decoding module. The goal of the decoder is to expand the encoded representation into a new object that is at the same level of abstraction of the model input. This is particularly common in sequence translation tasks, but is not required for embedding learning.
In addition, positional encodings are not used, and the original encoding blocks are replaced by multi-head GAT blocks. These changes make the whole model equivalent to a generalised graph neural network, meaning that the model maps token embeddings according to: \(\mathbf e_{q_i}\mapsto\phi\Big(\mathbf e_{q_i}, \bigoplus_{j\in\mathcal{N}_i} a(\mathbf e_{q_i},\mathbf e_{q_j})\psi(\mathbf e_{q_j})\Big)\), where \(\mathcal N_i\) is the neighborhood of node \(i\), \(\bigoplus\) is a permutation-invariant aggregator, \(a(\cdot,\cdot)\) is the attention scoring function, \(\psi\) is a local node function and \(\phi\) is a permutation invariant function.

Therefore, unlike the original Transformer, the proposed model is equivariant to permutations to the tokens in the input sequence: \(\mathbf Pf(\mathbf e, A)=f(\mathbf P\mathbf e, \mathbf PA\mathbf P^\mathsf T)\), where \(\mathbf P\) is an \(L\times L\) permutation matrix, \(A\) is the adjacency matrix of the query topology graph, and \(f\) represents the application of the model to the input sequence embeddings \(\mathbf e\) with query topology \(A\).
The main advantage in a network cybersecurity setting is increased robustness to small network timing perturbations, as a consequence of the final representation of each token being independent of its position within the input sequence.
This property also holds in the case of multiple adjacency matrices.



The main model component is a multi-head graph attention network, and a schematic figure is provided in our external Appendix: \url{https://github.com/m-altieri/DNS-GT}.
Analogously to \cite{Vaswani2017Attention}, we use \(\eta\) attention heads, that learn different linear projections for the query, key and value, projecting the embeddings with \(W_i^Q,W_i^K,W_i^V\in\mathbb{R}^{N\times (N/\eta)}\), for embedding vectors of size \(N\).
Given embeddings \(\mathbf e\), the attention output of each head is given by the scaled dot-product attention:

\begin{align}
\mathbf{S}_i &= \frac{\mathbf eW_i^Q(\mathbf eW_i^K)^\mathsf T}{\sqrt{N/\eta}} \\
\text{head}_i(\mathbf e, A) &= \text{MaskedSoftmax}(\mathbf{S}_i, A)\mathbf eW_i^V
\end{align}

and their results are concatenated and linearly projected with:
$$\text{Attn}(\mathbf e, A)=\big(\text{head}_1(\mathbf e, A)||\ldots||\text{head}_\eta(\mathbf e, A)\big)W^O,$$
mapping back to the original dimension with \(W^O\in\mathbb{R}^{N\times N}\).
The second argument of the MaskedSoftmax function is used as an element-wise mask to exclude elements from the softmax calculation corresponding to zeros in \(A\).
The factor \(A\in\mathbb{R}^{L\times L}\) is the adjacency matrix representing the sequence topology, as described in Section \ref{sec:topologies}, and is used to enable hard~\cite{bahdanau2014neural} attention,
which allows queries to only attend to keys of connected nodes in the graph.
This way, nodes that are not in the neighborhood of the current node are filtered out and their respective attention scores will be \(0\).
This matrix is used, for instance, to exclude \texttt{<PAD>} tokens from the attention mechanism, so that their embedding cannot affect the other token embeddings in any way, by setting to \(0\) the edges connecting \texttt{<PAD>} to any other token.
The multi-head graph attention is performed separately for each one of such matrices and results are aggregated: \(\bigoplus_{A\in\mathcal A} \text{Attn}(\mathbf e,A)\).

After the multi-head GAT computation, the result is added to the initial representation for the current block by means of a residual connection, and the result is normalised. 
This context-aware token representation is projected using a neural network with a nonlinear activation.
In particular, as in~\cite{Vaswani2017Attention}, two linear layers are used with a nonlinear layer in between: $$\max\left(0, \left(\mathbf x\mathbf W_1+\mathbf b_1\right)\mathbf W_2+\mathbf b_2\right)\mathbf W_3+\mathbf b_3,$$ where, for embeddings of size \(N\), \(\mathbf W_1,\mathbf W_3\in\mathbb{R}^{N\times N}\) and \(\mathbf W_2\in\mathbb{R}^{N\times 4N}\) are weight matrices, and \(\mathbf b_1,\mathbf b_2,\mathbf b_3\) are bias vectors.
Finally, we adopt another residual connection to add together the output of this neural network with the output of the multi-head GAT module, and the result is normalised to constitute the final output of the block.
The output embeddings of the whole block are recursively used as input for another identical block.

The idea is that each additional block propagates query representations towards the adjacent queries in the sequence. Therefore, each model block corresponds to a hop in the query adjacency graphs.

The final representation is linearly mapped with a softmax activation function, to obtain the probability for each domain in the vocabulary of appearing in place of the masked tokens.
The loss is computed only for masked tokens as the cross entropy between the model prediction and the ground truth.

\subsection{Applications}
\label{sec:applications}

The learned domain name embeddings enable two main types of downstream applications.
The first treats the embeddings learned at training time as standalone feature vectors for each domain. In this setup, the model serves as a feature extractor, and the embeddings are used with other models, for classification, clustering, anomaly detection, or other tasks. Any model that accepts vector inputs, from linear regression to neural networks, can be applied. However, the representations learned with this approach do not capture contextual information from surrounding domains.

The second type leverages the model’s architecture and hidden layers to incorporate context from domain sequences via the attention mechanism. A domain may carry different meanings depending on its sequence, enabling tasks like session classification. This approach requires using the trained model itself, following a ULMFiT-like procedure~\cite{howard-ruder-2018-universal}, where masked-token training serves as pre-training, and a classification head is added and trained for the final task, with model weights fully or partially frozen.

In practice, this method could be integrated into DNS resolvers to detect user behavior patterns in real time with contextual awareness. For example, it could block access to malicious domains or identify fraudulent users (e.g., botnets) by analyzing DNS traffic.

The method can detect behaviors involving suspicious domains, such as botnets, DNS tunneling, or access to risky sites, but generally only at the level of individual users, limiting its ability to detect multi-host attacks like fast-flux.
Notably, this approach is not limited to malicious domain detection. It can be extended to various downstream tasks, including host- and query-level ones, and is not restricted to classification.

\section{Experiments}
\label{sec:experiments}

\subsection{Dataset}
\label{sec:dataset}

The dataset used for our experiments is based on the TI-2016~\cite{9ync-vv09-19} dataset, which contains ten days of real DNS traffic collected from a campus network consisting of 4000 hosts during peak-load hours in the PCAP format. 
Specifically, the data in the original dataset amounts to $\approx\,$127 million DNS requests and $\approx\,$169 million DNS responses, for a total of $\approx\,$296 million DNS packets (for a total of 85.49 GB). 


For the training phase of the model, the first seven days were used, accounting for a total of \(\approx\! 13\) million queries. The remaining 3 days of network activities are used for evaluation, for a total of \(\approx\! 3.1\) million queries. Overall, this constitutes about $\approx\,$6.4\% of the original data. 

For the domain name classification task, we extracted and concatenated domain names from a public blacklist\footnote{https://firebog.net/}, identifying \(169\,066\) blacklisted domains within the dataset. These include activities such as privacy violations, user tracking without consent, and personal data collection, which we collectively refer to as malicious. 

\subsection{Setup}

Our method requires a pre-training stage, as described in Section~\ref{sec:method}, with the same experimental setup used for both pre-training and domain names classification.
The model is pre-trained on sequences of length \(L=32\), grouped by host IP, with host information merged using \(\omega=1\).
The domain vocabulary is trimmed to the \(30\,000\) most frequent domains, to exclude rare queries.
This size aligns well with common NLP vocabularies (e.g., BERT~\cite{devlin2018bert}). 
For the model architecture, we use 256-dimensional embeddings, 8 multi-head GAT blocks, and 8 attention heads.
The masking probability is set to \(10\%\), where masked domain are replaced with a \texttt{<MASK>} token \(90\%\), a random domain name \(10\%\), or is left unchanged \(10\%\), following standard practice~\cite{Vaswani2017Attention}.
Training was carried out for 1 million steps using a learning rate of \(0.0001\) and batch size of \(256\), running on two NVIDIA RTX A6000 GPUs.

We reserve $20\%$ of domains for evaluation and train on the remaining $80\%$.
Classification error is computed only on test set domains.

We compare our model against Word2Vec, a popular language model, for domain name embedding learning using DNS queries. 

%
For the evaluation, the DNS traffic is split into 7 days for training and 3 for testing.
A 5-fold cross validation is applied over the domain vocabulary: pre-training uses all domains, classification uses 80\% of domains for training and tests on the remaining 20\%, using blacklist labels as ground truth.
%
Results for Word2Vec-CBOW, Word2Vec-SkipGram, and our DNS-GT are reported in Table \ref{tab:classification} (ROC AUC, F1-score at 0.5 and optimal threshold), with ROC curves shown in Figure \ref{fig:roc}.



\subsection{Results and Discussion}
This subsection presents the quantitative results of our study. First, we analyze experimental external classifier performance, followed by end-to-end classification results (including ROC-AUC), and finally we discuss computational complexity.

\vspace{3pt}
\noindent\textbf{External classifiers:}
We quantitatively evaluate the classification of the blacklisted domain names using the dataset from Section~\ref{sec:dataset}, comparing the two modalities described in Section \ref{sec:applications}: \textit{i)} external classifiers, and \textit{ii)} end-to-end classification. 
For the external classifiers, we use AdaBoost, Decision Tree, Gaussian Na\"ive Bayes, SVM, Variational Auto-Encoder (VAE), Angle-base Outlier Detection (ABOD), Histogram-based Outlier Detection (HBOS), and Scalable Unsupervised Outlier Detection (SUOD). 
These models rely solely on the learned embeddings as input features, without context from surrounding queries.

\begin{table*}[]
\small
    \caption{Performance results of the considered methods on the domain names classification task, measured on the ROC AUC, F1 score with threshold $0.5$, and F1 score with the best threshold, for the three sequencing approaches.
    Bold text highlights the best results for each metric and each sequencing approach. The best results for each metric are surrounded by a rectangle.}
    \centering
    \resizebox{\linewidth}{!}{
    \begin{tabular}{lccccccccc}
        \toprule
        \multirow{2}[2]{*}{Models (w/ External Classifiers)} & \multicolumn{3}{c}{Fixed} & \multicolumn{3}{c}{Time} & \multicolumn{3}{c}{Density} \\
        \cmidrule(lr){2-4} \cmidrule(lr){5-7} \cmidrule(lr){8-10}
        & AUC & F1 @ 0.5 & F1 @ Best & AUC & F1 @ 0.5 & F1 @ Best & AUC & F1 @ 0.5 & F1 @ Best \\
         \midrule
         W2V-CBOW + AdaBoost &
         0.790 & 0.448 & 0.448 & 0.722 & 0.396 & 0.396 & 0.748 & 0.411 & 0.411 \\
         W2V-CBOW + Decision Tree &
         0.620 & 0.348 & 0.348 & 0.590 & 0.320 & 0.320 & 0.606 & 0.335 & 0.335 \\
         W2V-CBOW + GNB &
         0.821 & 0.494 & 0.508 & 0.678 & 0.348 & 0.356 & 0.738 & 0.427 & 0.433 \\
         W2V-CBOW + SVM &
         \textbf{0.846} & \textbf{0.524} & \textbf{0.531} & 0.799 & 0.457 & 0.479 & \textbf{0.811} & \textbf{0.492} & \textbf{0.535} \\
         W2V-CBOW + VAE &
         0.600 & 0.110 & 0.318 & 0.581 & 0.076 & 0.316 & 0.598 & 0.144 & 0.305 \\
         W2V-CBOW + ABOD &
         0.643 & 0.281 & 0.281 & 0.614 & 0.282 & 0.306 & 0.605 & 0.282 & 0.294 \\
         W2V-CBOW + HBOS &
         0.595 & 0.102 & 0.316 & 0.572 & 0.053 & 0.316 & 0.596 & 0.129 & 0.302 \\
         W2V-CBOW + SUOD &
         0.609 & 0.056 & 0.321 & 0.576 & 0.078 & 0.320 & 0.616 & 0.106 & 0.313 \\
         \midrule
         W2V-SkipGram + AdaBoost &
         0.750 & 0.404 & 0.404 & 0.762 & 0.431 & 0.431 & 0.666 & 0.355 & 0.355 \\
         W2V-SkipGram + Decision Tree & 
         0.640 & 0.368 & 0.368 & 0.677 & 0.409 & 0.409 & 0.600 & 0.329 & 0.329 \\
         W2V-SkipGram + GNB &
         0.755 & 0.464 & 0.464 & \textbf{0.813} & \textbf{0.503} & 0.507 & 0.766 & 0.482 & 0.483 \\
         W2V-SkipGram + SVM & 
         0.781 & 0.489 & 0.489 & 0.805 & 0.491 & \textbf{0.510} & 0.722 & 0.374 & 0.409 \\
         W2V-SkipGram + VAE &
         0.599 & 0.167 & 0.306 & 0.600 & 0.110 & 0.318 & 0.542 & 0.090 & 0.293 \\
         W2V-SkipGram + ABOD &
         0.676 & 0.282 & 0.312 & 0.643 & 0.281 & 0.290 & 0.589 & 0.281 & 0.297 \\
         W2V-SkipGram + HBOS &
         0.596 & 0.230 & 0.305 & 0.595 & 0.102 & 0.316 & 0.534 & 0.071 & 0.296 \\
         W2V-SkipGram + SUOD &
         0.606 & 0.066 & 0.309 & 0.609 & 0.056 & 0.321 & 0.549 & 0.035 & 0.297 \\
         \midrule
         DNS-GT + AdaBoost & 
         0.657 & 0.340 & 0.340 & 0.698 & 0.374 & 0.374 & 0.722 & 0.395 & 0.395 \\
         DNS-GT + Decision Tree & 
         0.549 & 0.283 & 0.283 & 0.595 & 0.325 & 0.325 & 0.568 & 0.300 & 0.300 \\
         DNS-GT + GNB & 
         0.715 & 0.390 & 0.394 & 0.774 & 0.453 & 0.466 & 0.767 & 0.457 & 0.463 \\
         DNS-GT + SVM & 
         0.693 & 0.374 & 0.378 & 0.765 & 0.450 & 0.450 & 0.784 & 0.467 & 0.478 \\
         DNS-GT + VAE &
         0.610 & 0.119 & 0.299 & 0.674 & 0.350 & 0.350 & 0.651 & 0.290 & 0.342 \\
         DNS-GT + ABOD &
         0.552 & 0.285 & 0.287 & 0.580 & 0.283 & 0.298 & 0.583 & 0.281 & 0.290 \\
         DNS-GT + HBOS &
         0.610 & 0.076 & 0.312 & 0.678 & 0.304 & 0.357 & 0.650 & 0.264 & 0.338 \\
         DNS-GT + SUOD &
         0.605 & 0.041 & 0.305 & 0.651 & 0.231 & 0.332 & 0.649 & 0.216 & 0.344 \\
         \midrule
         \multirow{2}[2]{*}{Models (End-to-End)} & \multicolumn{3}{c}{Fixed} & \multicolumn{3}{c}{Time} & \multicolumn{3}{c}{Density} \\
        \cmidrule(lr){2-4} \cmidrule(lr){5-7} \cmidrule(lr){8-10}
        & AUC & F1 @ 0.5 & F1 @ Best & AUC & F1 @ 0.5 & F1 @ Best & AUC & F1 @ 0.5 & F1 @ Best \\ \midrule
         W2V-CBOW (End-to-End) & 
         0.591 & 0.326 & 0.351 & 0.579 & 0.284 & 0.309 & 0.779 & 0.458 & 0.596 \\
         W2V-SkipGram (End-to-End) & 
         0.682 & \textbf{0.427} & \textbf{0.427} & 0.688 & \textbf{0.408} & 0.408 & 0.656 & 0.269 & 0.496 \\
         DNS-GT (End-to-End) & 
         \textbf{0.716} & 0.379 & \textbf{0.427} & \textbf{0.709} & 0.339 & \textbf{0.465} & \fbox{\textbf{0.848}} & \fbox{\textbf{0.567}} & \fbox{\textbf{0.654}} \\
         \bottomrule
    \end{tabular}
    }
    \label{tab:classification}
\end{table*}

As shown in Table~\ref{tab:classification}, the best AUC and F1 @ 0.5 scores are achieved by W2V-CBOW + SVM in the Fixed ($0.846$ and $0.531$) and Density ($0.811$ and $0.492$) strategies, and by W2V-SkipGram + GNB in the Time strategy ($0.813$ and $0.503$).
Performance across configurations varies significantly: AUC values range from $0.595$ to $0.846$ (Fixed), from $0.576$ to $0.799$ (Time), and from $0.596$ to $0.811$ (Density). 
W2V-CBOW appears more robust than W2V-SkipGram in the Fixed (outperforming in 5 out of 8 configurations) and Density (6 out of 8 configurations) strategies, whereas W2V-SkipGram is superior in the Time setting (outperforming in all 8 configurations). 
DNS-GT performs poorly with external classifiers, likely due to the inability of simple classifiers to leverage the contextual information encoded in the embeddings. However, DNS-GT fully exploits contextual information of related queries, resulting in a proper characterization of the network activity when the learned embeddings are synergically used together. Considering that base classifiers use domain embeddings in isolation, this discrepancy may result in a decrease in model performance. 
%

\vspace{3pt}
\noindent\textbf{End-to-end-classification:}
In the end-to-end setting, training resumes from the saved pre-trained model, which was trained on the MLM task.
The final Softmax layer is replaced by a dropout layer (rate \(0.2\)), followed by a single dense neuron with sigmoid activation.
The model is then fine-tuned on the training set for binary domain classification, using binary cross-entropy loss.
We do not freeze any layer, as this empirically yielded better results.
Results in Table~\ref{tab:classification} show that W2V-CBOW performs worst in the Fixed and Time strategies, with AUC of $0.591$ and $0.579$, respectively. W2V-SkipGram performs better in both, achieving an AUC of $0.682$ (Fixed) and $0.688$ (Time).
In the Density strategy, instead, W2V-CBOW outperforms W2V-SkipGram with an AUC of $0.779$.
However, DNS-GT consistently achieves the best results across all strategies, with AUC of $0.716$ (Fixed), $0.709$ (Time), and $0.848$ (Density), outperforming all baselines in each case. Additionally, DNS-GT achieves the highest F1 @ Best across all setups.
Compared to external classifiers, DNS-GT shows improved performance in the end-to-end setting. Word2Vec-based models, by contrast, show comparable results across both modalities. While W2V-CBOW + SVM reaches an AUC of $0.846$—close to DNS-GT’s best AUC—its F1 @ Best is only $0.531$, compared to DNS-GT’s $0.654$ in the end-to-end Density strategy.
%
Overall, DNS-GT delivers the highest performance across all metrics and sequencing strategies (marked with a surrounding rectangle in Table~\ref{tab:classification}), thanks to its ability to capture rich contextual dependencies via graph attention blocks.

\vspace{3pt}
\textbf{Ablation study:}
To assess the contribution of different components in the proposed model, we conducted an ablation study by fine-tuning the pre-trained model under three configurations: \textit{(i)} without the attention mechanism, \textit{(ii)} without host information (by setting the host weighting coefficient $\omega=1.0$), and \textit{(iii)} the full model, including both attention and host information.
The results, reported in Table \ref{tab:ablation}, show that the attention mechanism plays a crucial role in the model performance. Removing it leads to a degradation of 0.438 in terms of AUC.
Host embeddings also contribute positibely, although to a lesser extent. Their removal results in a degradation of 0.086 in terms of AUC.
Ultimately, our study shows that both components contribute positively the model effectiveness.

\begin{figure}
    \centering
\includegraphics[trim=20 20 20 40,clip,width=0.99\linewidth]{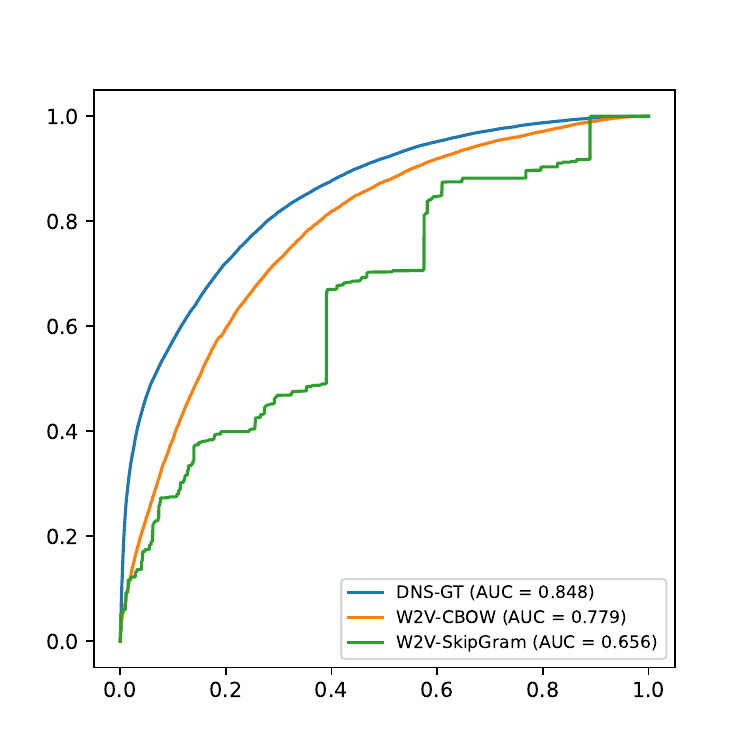}
    \caption{ROC curves for all considered methods with end-to-end training and evaluation.}
    \label{fig:roc}
\end{figure}

\vspace{3pt}
\textbf{ROC-AUC Analysis:}
The superior performance of DNS-GT in terms of ROC-AUC is further emphasized in Figure~\ref{fig:roc}, where its end-to-end variant consistently outperforms both baseline models across all false positive rates (x-axis).

\begin{table}[]
\small
    \centering
    \resizebox{\linewidth}{!}{
        \begin{tabular}{lccc}
            \toprule
             Configuration & AUC & F1 @ 0.5 & F1 @ Best \\ \midrule
             DNS-GT (No Attention) & 
             0.410 & 0.313 & 0.331 \\
             DNS-GT (No Hosts) & 
             0.762 & 0.467 & 0.508 \\
             DNS-GT (Complete) & 
             \textbf{0.848} & \textbf{0.567} & \textbf{0.654} \\
             \bottomrule
        \end{tabular}
    }
    \caption{Ablation study on the domain name classification task, comparing the proposed model in its full configuration with two variants where key components are deactivated.} 
    \label{tab:ablation}
\end{table}

\begin{table}[]
\small
    \centering
    \resizebox{\linewidth}{!}{
        \begin{tabular}{lccc}
            \toprule
             Models & Accuracy & AUC & F1 \\ \midrule
             W2V-CBOW (End-to-End) & 
             0.851 & 0.967 & 0.851 \\
             W2V-SkipGram (End-to-End) & 
             0.877 & 0.970 & 0.877 \\
             DNS-GT (End-to-End) & 
             0.877 & 0.970 & 0.877 \\
             \bottomrule
        \end{tabular}
    }
    \caption{Results of the considered methods on the botnet detection task, measured in terms of Accuracy, AUC, and F1 score (density-based sequencing approach).}
    \label{tab:botnet}
\end{table}

\vspace{3pt}
\textbf{Botnet detection:}
Botnets are networks of compromised machines controlled remotely by an attacker. They pose a significant challenge in cybersecurity due to the dynamic nature of host behaviour, as benign hosts can become infected at any time.
Using the same dataset and pre-training setup as previous experiments, we fine-tuned our model on the botnet detection task, where the objective is to classify hosts into categories such as clean, unknown, modpack, virut, necurs, conficker, or pitou (labels are sourced from the original dataset).
As shown in Table \ref{tab:botnet}, DNS-GT outperforms W2V-CBOW and performs on par with W2V-SkipGram. This result is somewhat unexpected given that W2V-SkipGram lacks strong contextual modeling capabilities. A possible explanation is that, during the pre-training stage, our model is incentivized to exploit contextual information in queries to reconstruct masked queries. During finetuning, the same mechanism for contextual extraction is not equally viable in determining the botnet status of a given host because other hosts do not provide relevant information.

%
Nevertheless, DNS-GT's advanced contextual modeling does not degrade accuracy in this task, matching the results of W2V-SkipGram. 
On the contrary, our proposed model significantly outperformed W2V-SkipGram in the domain classification task, where it is able to exploit contextual signals from diverse domain sequences.

\vspace{3pt}
\textbf{Computational complexity:}
We also assess the computational complexity of each method by measuring training and fine-tuning times, reported in Table~\ref{tab:time-complexity}.
As expected, DNS-GT presents the highest execution time, except during finetuning with the Density strategy, due to its more complex architecture and higher parameter count (24M) compared to W2V-CBOW and W2V-SkipGram (15M).
Training time is also strongly influenced by the sequencing strategy: the Density strategy is faster by about an order of magnitude, while the other two strategies exhibit similar time duration.
In summary, DNS-GT achieves higher accuracy at the expense of increased computational demand, especially in the end-to-end modality and with the Density strategy.

Additional performance metrics, such as latency time, cold start, and throughput, as well as other analyses involving the embedding space, are included in the Appendix.

\begin{table}[]
\small
\centering
\label{tab:time-complexity}
\resizebox{\linewidth}{!}{
    \begin{tabular}{llcc}
        \toprule
        \multirow{2}[2]{*}{Model} & \multirow{2}[2]{*}{Seq. strategy} & \multicolumn{2}{c}{Training time (minutes)} \\ \cmidrule(lr){3-4}
        && Pre-training & Finetuning \\
        \midrule
        \multirow{3}{*}{W2V-CBOW} & Fixed & 224 & 412 \\
        & Time & 204 & 401 \\
        & Density & 45 & 99 \\ \cmidrule(lr){1-4}
        \multirow{3}{*}{W2V-SkipGram} & Fixed & 482 & 529 \\ 
        & Time & 224 & 355 \\
        & Density & 47 & 107 \\ \cmidrule(lr){1-4}
        \multirow{3}{*}{DNS-GT} & Fixed & 1,933 & 1,009 \\
        & Time & 2,088 & 1,001 \\
        & Density & 116 & 62 \\
        \bottomrule
    \end{tabular}
}
\caption{Empirical analysis of time complexity for all considered models on end-to-end training.}
\label{tab:times}
\end{table}


%




\section{Conclusion}
\label{sec:conclusion}

In this paper, we presented a novel approach to network intrusion detection and prevention through the use of large-scale language models and graph neural networks applied to DNS traffic. Our proposed architecture leverages attention mechanisms and graph-based modeling to enable context-aware representations of domain names, learning an embedding space from DNS queries in an unsupervised manner.
%
We demonstrated that this approach not only captures the semantics of DNS traffic but also generalizes effectively to downstream cybersecurity tasks. Specifically, our analysis of the learned embedding space revealed a clear separation between benign and malicious domains, and our quantitative experiments on domain classification showed consistent improvements over baseline models. These results, together with several focused analyses, suggest that the model is successfully leveraging contextual information across query sequences, rather than treating domain names in isolation.
%
Our findings highlight the potential of building foundation models for cybersecurity applications based on DNS traffic. While further validation is required, this work opens the path to robust, scalable, and general language models for NIDP systems.

Looking forward, several promising research directions emerge:
First, a number of relevant methodological developments introduced in this work require deeper investigation, such as the integration of graph-based knowledge and the shared learning of hosts and domain name embeddings. The applicability of our model could also be explored using different types of data sources, to better assess its potential as a foundation model for NIDPS applications.
Second, a wider range of downstream tasks could be studied, such as session classification or multi-class domain characterization, to further demonstrate the versatility of our approach in different cybersecurity contexts. 
Third, building on our promising results, future work may investitgate the scalability of our model with larger datasets, under the hypothesis that general scaling laws might apply in this domain as well.
%
Finally, a comparison with other NIDPS techniques on specific network intrusion tasks based on DNS traffic could be pursued, although, to the best of our knowledge, no comprehensive public benchmark currently exists for such comparisons.


\appendix

\section{Preliminaries}
\label{sect:background}
%
The main analogies between our domain and NLP are described in Table~\ref{tab:mapping_nlp_dns} outlines a mapping for some of these concepts.
\begin{table}[htp]
    \centering

    \begin{tabular}{lll}
        \toprule
        \textbf{Concepts} & \textbf{NLP} & \textbf{DNS} \\ \midrule
        Token & Word$^a$ & Domain name \\
        Sequence & Sentence & Set of queries \\
        Order matters & Yes & No \\
        Semantics & Meaning &  User activity \\
        \bottomrule
    \end{tabular}
    \caption{\label{tab:mapping_nlp_dns}Analogies between NLP and DNS concepts.}
\end{table}

\paragraph{Pre-training Stage}
Pre-training is a preliminary phase of Large Language Model (LLM) development where the model learns general language representations from a large text corpus without specific task supervision. This phase employs self-supervised objectives that derive supervision signals from the data itself, eliminating the need for manual labeling.
There exist several pre-training strategies.
In this paper, we employ Masked Language Modeling (MLM) because it does not impose an ordering on the tokens. MLM involves randomly masking tokens in the input and training the model to reconstruct them. This bidirectional approach allows models to develop representations that incorporate both the left and right contexts. Formally, given an input sequence $\mathbf x=[q_1,q_2,\ldots,q_l]$, a subset of tokens
$M$ is masked, and the model is trained to predict these masked tokens:
$$\mathcal L_{MLM}(\theta)=-\mathbb E_{\mathbf x\sim{\mathcal D}}\sum_{i\in M} \log P_\theta (q_i|\mathbf x_{\setminus M}),$$
where $\mathbf x_{\setminus M}$ represents the sequence with masked tokens at random positions.
In the specific pre-training of our method, the objective is to correctly reconstruct masked domain names rather than entire queries. This is motivated by the sequences being composed of queries requested by a single host. 
The standard implementation masks approximately 15\% of the input tokens, replacing them with a special [MASK] token with probability 80\%, a random token with probability 10\%, and with the original token (unchanged) with probability 10\%.
This design mitigates the discrepancy between pre-training (with mask tokens) and fine-tuning (without mask tokens). The randomization strategy forces the model to maintain a strong representation for each input token.

In our workflow, the ultimate goal of the pre-training phase is to use unlabeled sequences $\mathbf x$ as inputs to learn a pre-trained model $\xi$. 

\paragraph{Finetuning and Classification Stages}
Finetuning adapts pre-trained models to specific downstream tasks using labeled data. This transfer learning approach leverages the general language understanding developed during pre-training and specializes it for particular applications.
For classification tasks, a task-specific head (typically a feed-forward neural network) is attached to the pre-trained model's output representations. The entire model with this additional head is then trained on labeled examples, with a task-specific loss function:
$$\mathcal L_{task}(\theta) = \mathcal L_{CE}(f_\theta(\mathbf x), y),$$
where $f_\theta$ represents the model with parameters $\theta$, mapping input $\mathbf x$ to predictions and $\mathcal L_{CE}$ is typically cross-entropy loss between predictions and ground truth $y$.

In our workflow, the ultimate goal of the fine-tuning phase is to leverage labeled input sequences $\mathbf{\bar x}$ and the pre-trained model to fine-tune the latter, resulting in an updated model $\xi'$. The fine-tuning process can be performed several times with the same base model, to specialize it into different downstream tasks. 
Once the model is fine-tuned, it can be used for inference to classify new, previously unseen sequences.

\paragraph{Model: Attention Mechanism}
Attention allows a model to weigh the importance of different tokens in a sequence when encoding each token. The standard implementation is called "scaled dot-product attention" \cite{Vaswani2017Attention} and proceeds as follows. For a given input sequence, three matrices are computed: queries ($Q$), representing what the current token is looking for, keys ($K$), representing what each token in the sequence offers, and values ($V$), representing the actual content of each token.
These are calculated by multiplying the input embeddings by the learned weight matrices $W^Q$, $W^K$ and $W^V$: $Q=XW^Q$, $K=XW^K$, $V=XW=V$, where $X\in\mathbb R^{n\times d}$ is the input matrix for a sequence of length $n$ with embedding dimension $d$.
The attention scores are then calculated as:
$$\text{Attn}(Q,K,V)=\text{softmax}\left(\frac{QK^T}{\sqrt{d_k}}\right)V.$$
The result is a new representation for each token that incorporates information from other relevant tokens in the sequence.
Rather than performing attention once, transformers implement multi-head attention, which applies the attention mechanism in parallel across different representation subspaces:
\begin{multline}
\text{MultiHead}(Q,K,V) = \\
\text{Concat}(\text{head}_1,\text{head}_2,\ldots,\text{head}_h)W^O
\end{multline}
where each head is:
$$\text{head}_i=\text{Attn}(QW_i^Q,KW_i^K,VW_i^V).$$
Each head uses different learned projections $W_i^Q$, $W_i^K$, and $W_i^V$, allowing the model to attend to information from different representatio subspaces. The outputs are concatenated and projected with $W^O$ to produce the final result.
This multi-head approach allows the model to simultaneously attend to different aspects of the input across different heads.

In our model, the attention mechanism takes place at the beginning of each Transformer block, as shown in Figure \ref{fig:blocks}.  

\paragraph{Model: Graph Attention Network Component}
Graph Attention Networks (GATs) represent a natural extension of the transformer attention mechanism to graph-structured data. While standard transformers operate on sequence data with an implicit fully-connected attention graph, GATs explicitly model and operate on arbitrary graph structures.
The fundamental operation in Graph Attention Networks parallels transformer attention but adapts it to graph topology. For a node $i$ with features $h_i$ in a graph:
$$h'_i = \sigma\left(\sum_{j\in\mathcal N_i}\alpha_{ij}Wh_j\right),$$
where $\mathcal N_i$ is the neighborhood of node $i$, $\alpha_{ij}$ are attention coefficients, $W$ is a learned weight matrix, $\alpha$ is a non-linear activation function.
Graph data typically lacks a natural ordering, so GATs do not use positional encodings. Instead, the graph structure itself implicitly encodes the relative positions.

In our model, the GAT mechanism is exploited to take into account domain topologies in the attention phase. 

\section{Data Preprocessing}
In this stage, raw DNS traffic data is processed to convert the PCAP format to ordered sequences. We filter a subset of relevant sequences based on a few filtering criteria (e.g., each sequence contains queries from the same host). 
As a result, a subset of the full dataset is retained, resulting in sequences $\mathbf x = [q_1,q_2,\ldots,q_l]$, where each $q_i$ is a pair $(h_i, d_i)$ made up of a host and a domain name.
The output of this step consists of two subsets of sequences: unlabeled sequences $\mathbf x$ (used in the pre-training phase) and labeled sequences $\mathbf{\bar x}$ (used in the fine-tuning phase). 
%

The initial data is a collection of PCAP files~\cite{pcap_spec} containing DNS network traffic collected from one or multiple DNS resolvers within a network.
The data pipeline aims at cleaning and preprocessing these files to produce suitable DNS queries for training, and is outlined by the following steps:
\begin{itemize}
\item \emph{filtering out unsuitable host IPs}, to include only queries that have been sent from IPs that are machines operated by active end-users. Specifically, host IPs are filtered out in any of these two cases: \textit{(i)} too low number of requests sent (in our specific case 100), indicating that the user is inactive \footnote{{Given the specific nature of the proposed model and learning process, although a number of requests sent being below a certain threshold does not necessarily imply  inactivity, a host with very few occurrences in the dataset would not lead to meaningful embeddings due to a lack of training examples.
Such instances would effectively constitute outliers, or noise, for learning purposes, and are therefore removed to improve the quality of final embeddings and the computational complexity of the method.}}, or \textit{(ii)} too high or too low ratio of requests to responses (in our case case higher than \(1.015\) or lower than \(0.985\)), indicating that the host may not be an end-user but rather a DNS resolver;
\item \emph{selecting only `A' records}, to focus only on requests sent with the goal to resolve a domain name into its corresponding IPv4 address. {Although other types of records, such as `RDATA', could be exploited for additional information, this would require learning additional embeddings just to extract the correct semantics from the records, resulting in a significant increase in model complexity. We recognize it as a potential direction for future exploration};
\item \emph{selecting only requests}, by checking that the destination port is \(53\), and ignoring responses;
\item \emph{filtering out re-transmitted requests}, to avoid duplicates in the dataset.
\end{itemize}

After this cleaning process, the following attributes are extracted for each query: timestamp, source host IP, and requested domain name.
Thus, we can represent a query as a triple \(q=(h, d, t)\), where \(h\) denotes its source host, \(d\) the requested domain, and \(t\) the timestamp.

Queries sent by the same host are grouped together, so that queries belonging to the same group are semantically related and present some underlying patterns that can be extracted. The choice of grouping by host allows to reduce the likelihood of noise by avoiding activities from different hosts (and therefore different users) to be mixed in the same sequence.
In the NLP analogy, this can be seen as the equivalent of grouping words gathered from the same text source when training NLP models, as opposed to taking words gathered from multiple different texts simultaneously. It is rather intuitive how using a single source at a time provides sequences that have a more cohesive internal semantics, and allows for the extraction of more meaningful patterns.

\begin{figure*}[htp]
    \centering
    \includegraphics[width=0.8\linewidth]{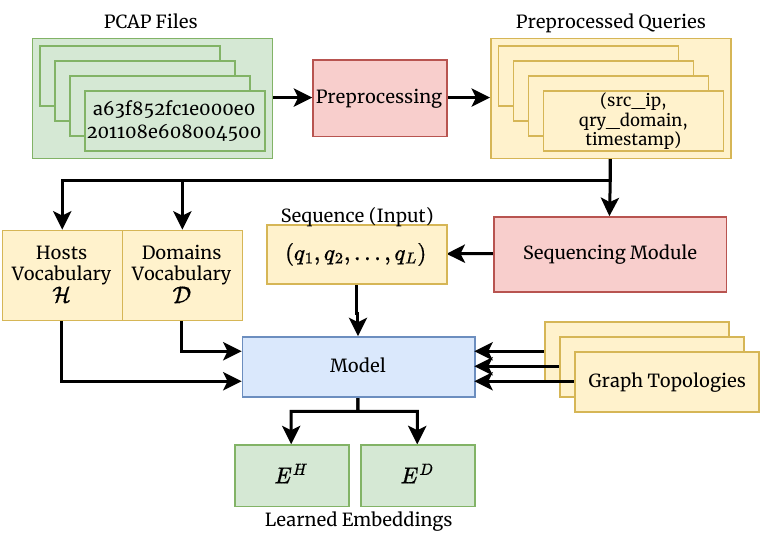}
    \caption{High-level overview of the training process and the data pipeline for the proposed methodology to learn representations of DNS queries. 
    The colour encodes the function of the stage: input and output (green), data (yellow), operations (red) and learnable neural networks (blue).}
    \label{fig:pipeline}
\end{figure*}

Figure~\ref{fig:pipeline} shows a graphical overview of the training process and the data pipeline.

\clearpage 

\begin{figure*}[htp]
    \centering
    \includegraphics[width=0.9\linewidth]{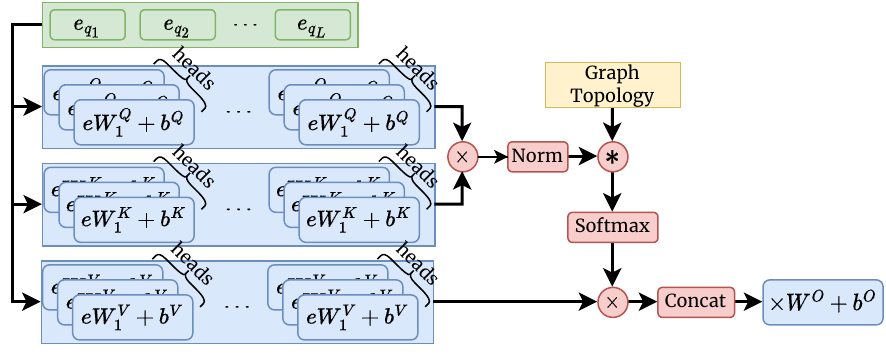}
    \caption{Architecture of the graph attention network for a single graph topology. The colour is used to emphasise input (green), tensors (yellow), operations (red) and learnable neural networks (blue).}
    \label{fig:gat}
\end{figure*}

Figure~\ref{fig:gat} shows a schematic view of the attention mechanism adopted in the architecture.

\begin{figure*}
    \centering
     \includegraphics[width=0.8\linewidth]{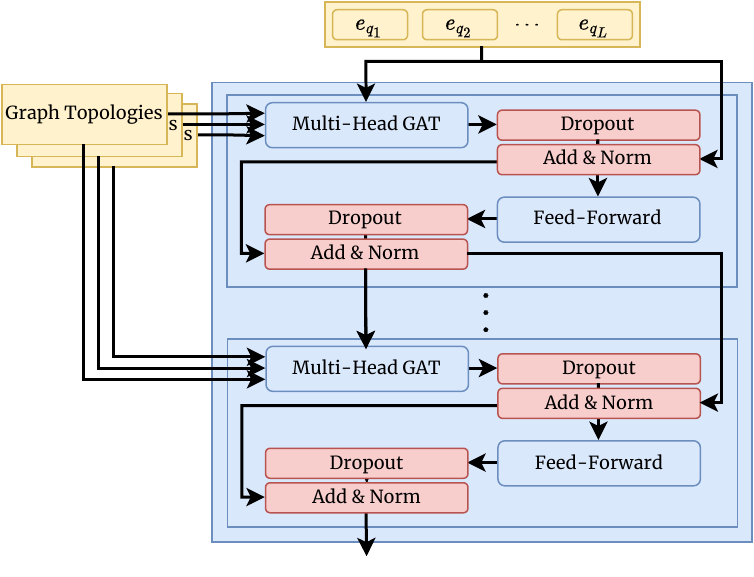}
    \caption{Stacking of multiple model blocks. The color is used to emphasise tensors (yellow), operations (red) and learnable neural networks (blue).}
    \label{fig:blocks}
\end{figure*}

Figure~\ref{fig:blocks} shows the stacking of multiple multi-head graph attention blocks.

\section{Graph Topologies}
\begin{figure*}[htp]
    \centering
    \includegraphics[width=0.7\linewidth]{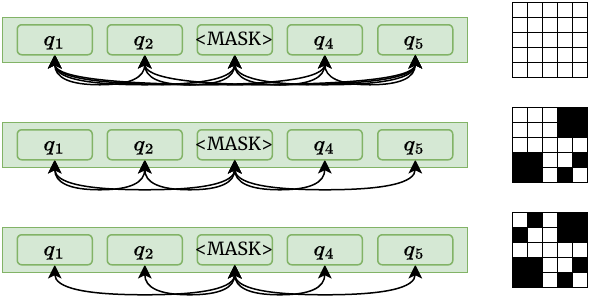}
    \caption{Examples of different graph topologies for a given sequence, with their corresponding adjacency matrices. Respectively, fully connected graph (top), a graph where the masked query is connected to all other queries and \(q_1\) and \(q_2\) are connected (middle), and a graph with where each node is only connected to the masked query (bottom). In these examples all nodes have self-loops.}
    \label{fig:topologies}
\end{figure*}

The way graph topologies are defined is context-specific, which is the reason why they are referred to as being knowledge-based.
Figure~\ref{fig:topologies} shows some basic examples of graph topologies in terms of token connections and their respective adjacency matrix.

\section{Baselines}

In our study, we consider Word2Vec as a baseline model for the classification task.
The main reason is its straightforward adaptability to different applications (in this particular context, by simply considering domain names as words in the vocabulary). 
More sophisticated alternatives such as BERT were not considered due to their reliance on the ordering of the elements of each sequence, and would require significant adaptations that are beyond the scope of this work.
It is important to note that although Word2Vec has been applied to DNS queries for embedding learning, its effectiveness for downstream tasks has not been studied. From this viewpoint, our experiments represent the first attempt to quantitatively measure and compare the effectiveness of different language models to learn domain names embeddings, evaluating them on the domain names classification task.

We consider both the Continuous Bag of Words (CBOW) and the Skip-Gram models.
In the former, we attempt to predict the current domain based on its neighboring domains, which represent the context.
In the latter, the goal is to predict neighboring domains based on the current domain.
The Word2Vec models are implemented to be efficient and in a way to have a comparison that is as fair as possible.
Specifically, the context for CBOW is computed in parallel for all tokens in the sequence, in an independent way.
In general, this is equivalent to 
$$\mathbf C=(\mathbf B_r-I_L) \cdot\underbrace{\begin{bmatrix}\text{---} & \mathbf e_1 & \text{---}\\\text{---} & \mathbf e_2 & \text{---}\\ & \dots & \\\text{---} & \mathbf e_L & \text{---}\end{bmatrix}}_{L\times N},$$
where the \(\mathbf e_i\)'s are the token embeddings for the given sequence, \(I_L\) is the identity matrix of size \(L\) and \(\mathbf B_r\) is a binary band matrix with bandwidth \(r\), and \(r\) corresponds to the context window size for the Word2Vec model.
In our case, we set \(r=L\) (and thus \(\mathbf B_r = \mathbf 1)\), so that all tokens in the sequence contribute to the context, which, for CBOW, is computed as the sum of the domain embeddings.
Skip-Gram, on the other hand, uses only the current domain embedding \(\mathbf e_i\) (without context) to predict the domains in the context window.
Moreover, we use the same sequence length \(L\) and embedding size \(N\) for Word2Vec and our proposed model.
With this implementation, the context \(\mathbf C\) is computed for all tokens at once, without the need to shift the window along the sequence.

In doing so, both models are able to learn vector representations for domain names, such that the similarity in the embedding space reflects the semantic similarity.
For the pre-training phase, the contextual embeddings are first projected to a hidden space, again in parallel: \(\mathbf H=\mathbf C\mathbf W_h+\mathbf b_h\), to obtain a representation that is used to predict the target token in the case of CBOW, or the other tokens in the sequence in the case of SkipGram: $$\mathbf z=\text{softmax}(\mathbf H\underbrace{\mathbf W_s}_{N\times V}+\mathbf b_s)$$

We adapt the learned Word2Vec models to support domain names classification. Specifically, we train an additional binary classification layer that we add on top of the model instead of the previous softmax layer, using the projected contextual representation as input, to produce a classification for all tokens in the sequence, with: 
$$\mathbf{\hat y}=\sigma(\mathbf H\underbrace{\mathbf W_c}_{N}+\mathbf b_c),$$
where \(\sigma\) is the sigmoid activation function.

\section{Intra- and inter-distance between clusters}

\begin{figure}
    \centering
    \includegraphics[width=\linewidth,trim=5 5 5 5,clip]{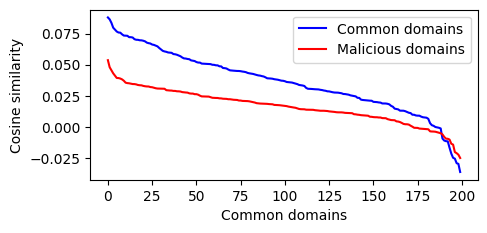} \\
    \ \ \ \ \ \ \ \ \ (a) \\
    ~ 
    \includegraphics[width=\linewidth,trim=5 5 0 0,clip]{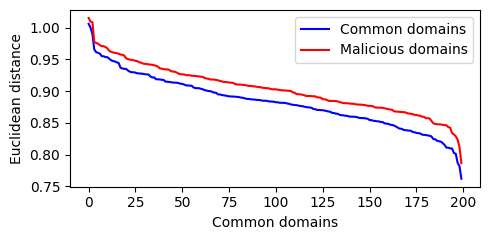} \\
    \ \ \ \ \ \ \ \ \ (b) \\
    ~ 
    \caption{Comparison of average intra-distance among the top 17 most common {benign} domains (blue line) and average inter-distance between common {benign} domains and blacklisted domains (red line) using the cosine similarity (a) and the Euclidean distance (b).}
    \label{fig:good_vs_mal}
\end{figure}

Using the same set of benign and blacklisted domain names, an evaluation of their embedding distance is conducted and is shown in Figure~\ref{fig:good_vs_mal}. In particular, the figure shows the average cosine similarity (a) and Euclidean distance (b) between common {benign} domains and other common {benign} domains (blue line), and the same metrics  between common {benign} domains and blacklisted domains (red line).
This visualization clearly shows that common {benign} domains tend to assume representations around the centre of the embedding space, and are closer to each other, while blacklisted domains tend to be scattered around the embedding space, further apart from each other and from common {benign} domains.



\begin{figure}
    \centering
    \includegraphics[width=\linewidth,trim=5 5 5 15,clip]{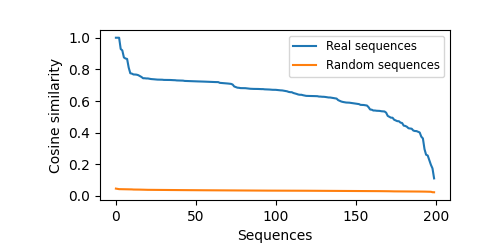}
    \includegraphics[width=\linewidth,trim=5 5 5 15,clip]{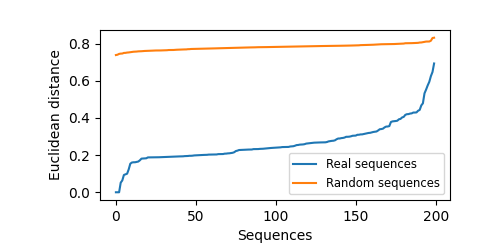}
    \caption{Average pairwise cosine similarity and Euclidean distance between domain name embeddings appearing in the same sequence vs. randomly sampled domains. 
    }
    \label{fig:real_vs_random_cos}
\end{figure}

    

In Figure~\ref{fig:real_vs_random_cos} 
we show the average cosine similarity and average Euclidean distance between embeddings of domain names the appear in the same sequence, for 200 randomly sampled sequences (blue line), and we compare it with the average cosine similarity and average Euclidean distance between embeddings of domain names that are randomly sampled from the vocabulary (orange line).
Each metric shows two comparisons: at the top, the comparison with the raw randomly sampled sequences, and at the bottom, the comparison with sequences where duplicated domains have been removed.
It is immediately apparent that the embeddings of domain names appearing in the same sequences are more similar (both in vector direction and Euclidean distance), which generally indicates that they are semantically related.

\section{Computational Complexity}
\begin{figure}
    \centering
\includegraphics[width=\linewidth]{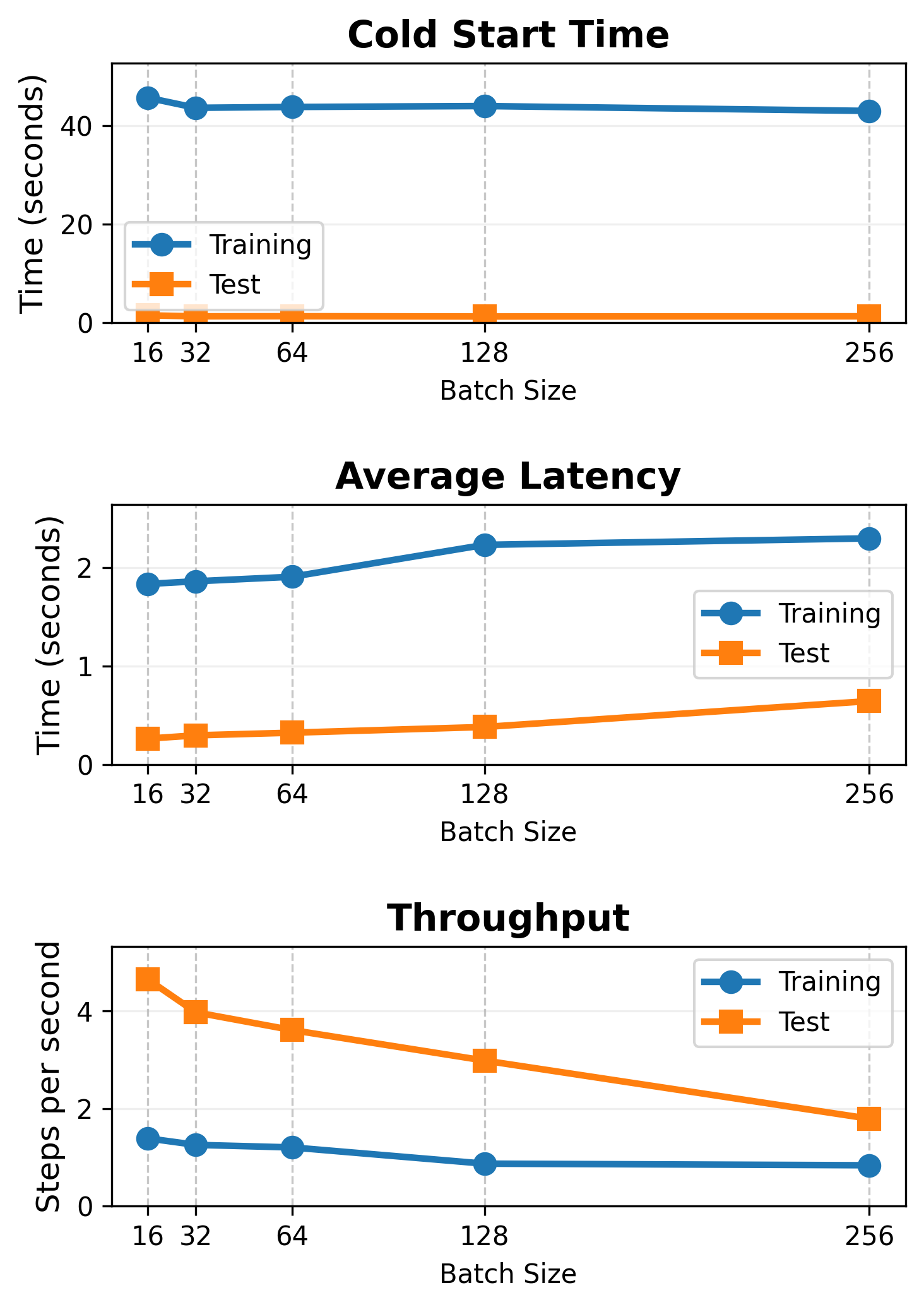}
    \caption{Cold start, latency and throughput of DNS-GT during both training and testing stages, with varying batch sizes, on the standard configuration used in the experiments. Steps per second corresponds to the number of batches processed in the time unit (second).}
    \label{fig:throughput}
\end{figure}

One useful perspective is provided by the average latency time, cold start, and throughput. 
After model initialization, cold start is defined as the time elapsed from the time the model is called until the first step (batch) is processed. 
Latency is the execution time required by the model to process a batch of a specific size, including cold start time, averaged across 50 executions.
Throughput is obtained as the number of steps (batches) processed in unit time (second). To obtain a realistic estimate representative of the working conditions of the model excluding external factors (such as in memory data loading, and other hardware implications), we skip the computation of throughput for the first 5 batches.
Results in Figure \ref{fig:throughput} show how latency, cold start, and throughput vary during both training and testing stages, considering increasing values of  batch size.
We observe that the cold start phase essentially requires the same amount of time regardless of batch size. Moreover, the latency time increases just minimally with increasing batch size. Finally, the throughput decreases sub-linearly with the batch size. Together, these results highlight that our model is scalable with respect to batch size and can be efficiently utilized in a real-world setting.

\section{Context sensitivity analysis}
We conduct additional experiments to ascertain that the model’s classification score is context-sensitive.

\begin{figure*}
    \centering
\includegraphics[width=\linewidth]{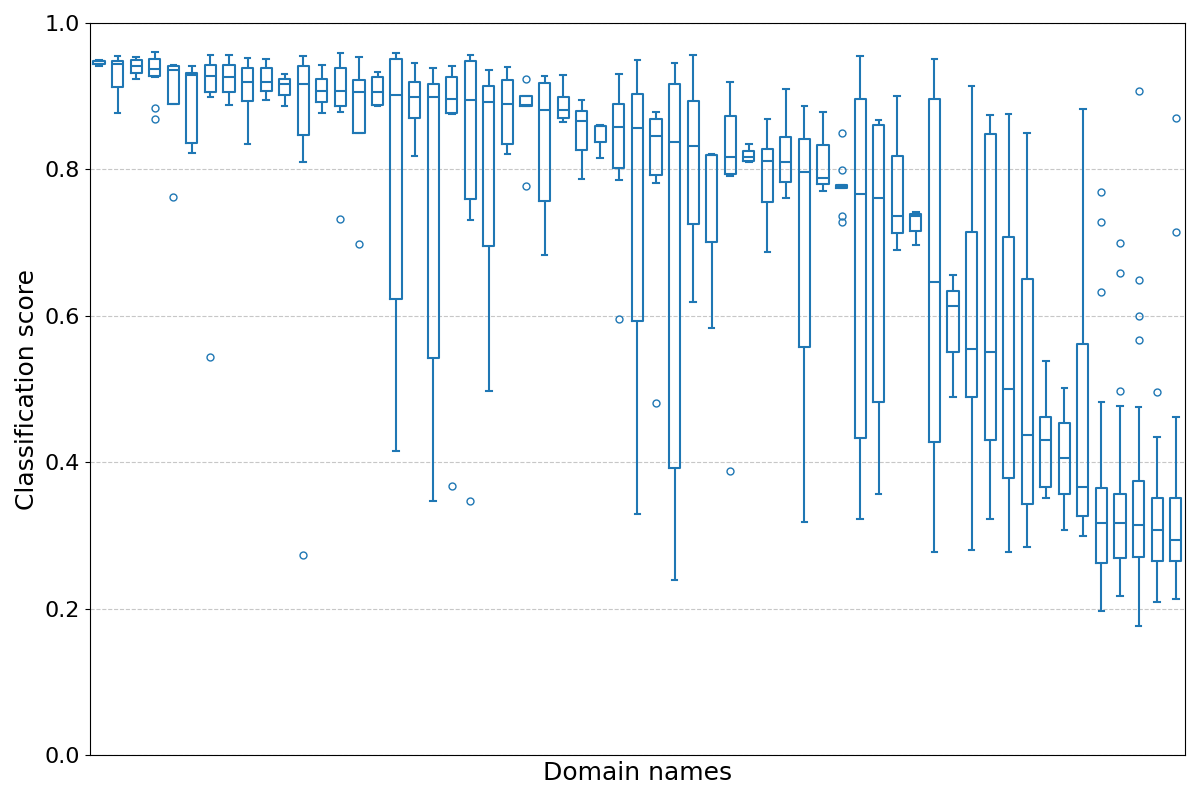}
    \caption{Box plot showing the classification scores for a random set of domain names across different sequences in which they appear. 
    For each domain name, the assigned classification scores vary according to the contextual information. In the plot, domains are sorted in descending order according to the median value of their classification scores.}
    \label{fig:boxplot}
\end{figure*}

First, we randomly select 200 domains. For each domain, we identify all sequences in which it appears. We require that each domain is present in at least 5 sequences. This filtering stage leads to a selection of 59 domains. Each sequence is then subject to model classification, and the variation in these scores is analyzed.
This information is represented as a box-plot visualization (see Figure \ref{fig:boxplot}). The visualization clearly shows that the model’s classification score varies significantly for the same domain across different sequences due to different contextual information.

Moreover, to statistically analyze the variation of classification scores, we resort to the Coefficient of Variation  as a standardized measure of dispersion of a probability distribution or frequency distribution. It is defined as the ratio of the standard deviation to the mean. This measure allows us to analyze a single domain with numerous repeated measures, i.e., classification score observed for different sequences/contexts.
For each domain, we compute the coefficient of variation on the list of classification scores (one entry for each sequence in which it appears) and verify whether this coefficient is above a threshold defined to represent a significant variation across scores. 

Across multiple executions, we observe that the variation of classification scores is significant for multiple threshold values. Specifically, the coefficient of variation exceeds 0.01 in $98\%$ of the cases, 0.03 in $88\%$ of the cases, and 0.05 in $75\%$ of the cases. This behavior denotes that the classification score is clearly influenced by the context in which the domain appears.  

The box plot in Figure \ref{fig:boxplot} shows the classification scores for a random set of domain names across different sequences in which they appear. This visualization highlights that the classification score of a domain presents a significant variation according to the contextual information.

\section{Inference example}

\begin{figure*}
\lstset{
    basicstyle=\ttfamily\tiny,
    keywordstyle=\bfseries,
    alsoletter={.-},
    morekeywords={scontent-hkg3-1.xx.fbcdn.net}
}
\begin{lstlisting}[basicstyle=\tiny,caption={Inference for a sample input sequence. The left part of each query is the model input \((h,d)\), and the right part is the model predicted domain for that position, with its probability.},captionpos=b]
172.31.1.4 lh5.ggpht.com ---------------------> lh5.ggpht.com (97.14%)
172.31.1.4 safebrowsing-cache.google.com -----> safebrowsing-cache.google.com (96.42%)
172.31.1.4 platform.twitter.com --------------> platform.twitter.com  (95.61%)
172.31.1.4 scontent.xx.fbcdn.net -------------> scontent.xx.fbcdn.net (98.63%)
172.31.1.4 fbcdn-photos-e-a.akamaihd.net -----> fbcdn-photos-e-a.akamaihd.net (96.77%)
172.31.1.4 vip.azurewebsites.windows.net -----> vip.azurewebsites.windows.net (42.97%)
172.31.1.4 go.microsoft.com ------------------> go.microsoft.com (98.23%)
172.31.1.4 <MASK> ----------------------------> scontent-hkg3-1.xx.fbcdn.net (63.46%)
172.31.1.4 r11---sn-5hne6nez.googlevideo.com -> r11---sn-5hne6nez.googlevideo.com (93.46%)
172.31.1.4 androidads21.adcolony.com ---------> androidads21.adcolony.com (99.81%)
172.31.1.4 android.googleapis.com ------------> android.googleapis.com (97.81%)
172.31.1.4 mob.d.aa.online-metrix.net --------> mob.d.aa.online-metrix.net (83.54%)
172.31.1.4 s3-1.amazonaws.com ----------------> s3-1.amazonaws.com (96.64%)
172.31.1.4 accounts.google.com ---------------> accounts.google.com (98.37%)
172.31.1.4 ws.mcafee.com ---------------------> ws.mcafee.com (99.32%)
...
\end{lstlisting}
\end{figure*}

The trained model is able to reconstruct masked queries only based on the other tokens in the sequences, which pass information through the graph attention layers. This information is combined in a useful way thanks to the learned hidden dependencies within training data.
A practical example of model inference is shown in Listing~1, with an input sequence for host \texttt{172.31.1.4}. The eighth domain in the sequence is masked, but the model is able to predict it correctly, assigning a \(63.46\%\) probability to the corresponding domain \texttt{scontent-hkg3-1.xx.fbcdn.net} in the vocabulary. This behaviour shows that our model is able to learn semantic information in a sequence of DNS queries, and is able to reconstruct missing parts of the sequence.

\section{Embeddings analysis}
\begin{figure}
    \centering
    \includegraphics[width=\linewidth]{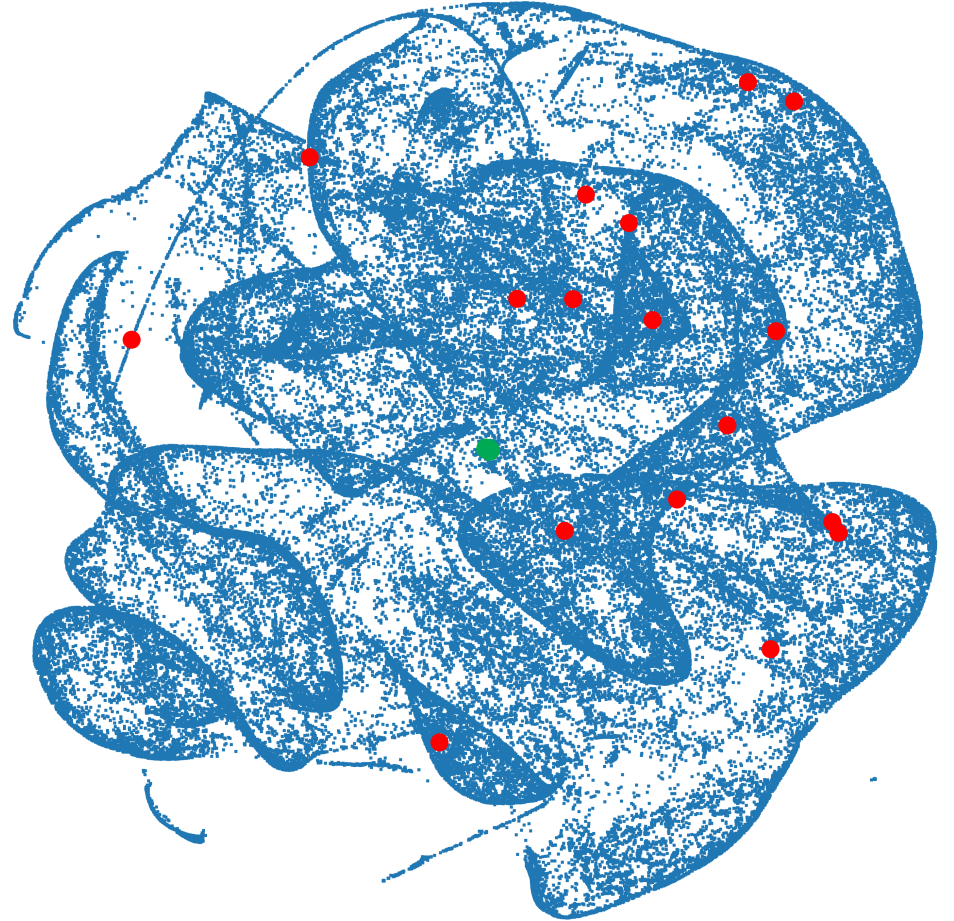}
    \caption{Embedding space visualization using t-SNE with perplexity 30 and 500 iterations. The red dots show the location of the blacklisted domain names, and the green dots (clustered at the very centre) represent the most common {benign} domain names in our dataset.}
    \label{fig:embeddings-space}
\end{figure}

A visualisation of the embedding space is conducted using the t-SNE~\cite{van2008visualizing} algorithm, by mapping multi-dimensional feature vectors (the learned embeddings) onto a 2D space. This allows us to visualise where the learned embeddings for common {benign} domain names are located in the embedding space, compared to blacklisted domain names. For this analysis, we used the KADhosts blacklist: \footnote{\url{https://raw.githubusercontent.com/PolishFiltersTeam/KADhosts/master/KADhosts.txt}}, which is one of the blacklists used as ground truth for the downstream task. This choice allows us to focus on a specific set of blacklisted domains, and to generate a human-interpretable visualization of the embedding space. Specifically, using this blacklist, we identify \(17\) matches for blacklisted domains in our dataset.

The learned domain name embedding space is displayed in Figure~\ref{fig:embeddings-space}. Interestingly, the figure highlights that benign domain names (marked in green) are clustered in the middle of the embedding space, whereas the 17 blacklisted domain names (marked in red) are spread across the plot, but are, in all cases, far away from the benign domains. 

\section{Context Extraction Capabilities}
As a further analysis, we investigate the ability of DNS-GT to extract and leverage contextual information during inference.
Specifically, we analyze cases in which a domain is generally benign, but may appear suspicious when observed within certain sequences.
Consider the domain \texttt{download.cdn.mozilla.net}, which is legitimately used by the Firefox browser to retrieve updates and related resources.
While this domain is typically benign, it may be exploited or spoofed for malicious purposes. Therefore, while being associated with a static label in the training data, DNS-GT is able to account for the surrounding context of each query, enabling dynamic interpretation during inference.
\begin{figure}

\lstset{
    basicstyle=\ttfamily\scriptsize,
    keywordstyle=\bfseries,
    alsoletter={.},
    morekeywords={download.cdn.mozilla.net}
}
\renewcommand{\ttdefault}{pcr}
\begin{lstlisting}[title={(a) No context}]
download.cdn.mozilla.net (0) -> 0.064
\end{lstlisting}

\begin{lstlisting}[title={(b) Benign context}]
download.cdn.mozilla.net (0) -> 0.023
clients3.google.com (0) ------> 0.019
edge-mqtt.facebook.com (0) ---> 0.024
mtalk4.google.com (0) --------> 0.023
graph.instagram.com (1) ------> 0.022
\end{lstlisting}

\begin{lstlisting}[title={(c) Suspicious context}]
download.cdn.mozilla.net (0) -> 0.854
us-u.openx.net (1) -----------> 0.782
cas.criteo.com (1) -----------> 0.826
lotame.nexac.com (1) ---------> 0.818
idsync.rlcdn.com (1) ---------> 0.838
su.addthis.com (1) -----------> 0.820
ps.eyeota.net (1) ------------> 0.688
lotame.nexac.com (1) ---------> 0.823
lotame.nexac.com (1) ---------> 0.744
lotame.nexac.com (1) ---------> 0.802
cat.hk.as.criteo.com (1) -----> 0.549
ip2.casalemedia.com (1) ------> 0.871
lotame.nexac.com (1) ---------> 0.658
lotame.nexac.com (1) ---------> 0.899
tag.contextweb.com (1) -------> 0.705
lotame.nexac.com (1) ---------> 0.726
\end{lstlisting}

\caption{Classification score of DNS-GT for \texttt{download.cdn.mozilla.net} in three different contexts. In (a) and (b), the context is either absent or benign, and the classification score is close to 0 (benign), while in (c), the context is suspicious, potentially indicating spoofing or other malicious activities, and the classification score is close to 1 (malicious).}
\label{fig:context}
\end{figure}
Figure~\ref{fig:context} showcases three scenarios involving \texttt{download.cdn.mozilla.net}: in (a), the domain appears in isolation and is correctly classified as benign; in (b), it occurs within a sequence of other benign queries and is again classified as benign; in (c), however, it occurs with domains commonly associated with advertisement, tracking, or audience segmentation services (domains that are labeled as `malicious' during training). In this last context, the model increases its suspicion and correctly classifies the domain name as malicious, possibly flagging a spoofed or anomalous use.
%
This context-sensitive behaviour is systematically observed across the dataset, as shown by another analysis in our external Appendix: \url{https://github.com/m-altieri/DNS-GT}. 

\bmhead{Authors’ Contributions}
Massimiliano Altieri: \textit{Conceptualization, Data Curation, Methodology, Investigation, Software, Visualization, Writing} -- Ronan Hamon: \textit{Conceptualization, Methodology, Data Curation, Writing} -- Roberto Corizzo: \textit{Investigation, Writing, Supervision} -- Michelangelo Ceci: \textit{Writing (Review \& Editing), Resources, Validation} -- Ignacio Sanchez: \textit{Writing (Review \& Editing), Resources, Supervision}

\bibliography{bibliography}


\begin{thebibliography}{39}
\ifx \bisbn   \undefined \def \bisbn  #1{ISBN #1}\fi
\ifx \binits  \undefined \def \binits#1{#1}\fi
\ifx \bauthor  \undefined \def \bauthor#1{#1}\fi
\ifx \batitle  \undefined \def \batitle#1{#1}\fi
\ifx \bjtitle  \undefined \def \bjtitle#1{#1}\fi
\ifx \bvolume  \undefined \def \bvolume#1{\textbf{#1}}\fi
\ifx \byear  \undefined \def \byear#1{#1}\fi
\ifx \bissue  \undefined \def \bissue#1{#1}\fi
\ifx \bfpage  \undefined \def \bfpage#1{#1}\fi
\ifx \blpage  \undefined \def \blpage #1{#1}\fi
\ifx \burl  \undefined \def \burl#1{\textsf{#1}}\fi
\ifx \doiurl  \undefined \def \doiurl#1{\url{https://doi.org/#1}}\fi
\ifx \betal  \undefined \def \betal{\textit{et al.}}\fi
\ifx \binstitute  \undefined \def \binstitute#1{#1}\fi
\ifx \binstitutionaled  \undefined \def \binstitutionaled#1{#1}\fi
\ifx \bctitle  \undefined \def \bctitle#1{#1}\fi
\ifx \beditor  \undefined \def \beditor#1{#1}\fi
\ifx \bpublisher  \undefined \def \bpublisher#1{#1}\fi
\ifx \bbtitle  \undefined \def \bbtitle#1{#1}\fi
\ifx \bedition  \undefined \def \bedition#1{#1}\fi
\ifx \bseriesno  \undefined \def \bseriesno#1{#1}\fi
\ifx \blocation  \undefined \def \blocation#1{#1}\fi
\ifx \bsertitle  \undefined \def \bsertitle#1{#1}\fi
\ifx \bsnm \undefined \def \bsnm#1{#1}\fi
\ifx \bsuffix \undefined \def \bsuffix#1{#1}\fi
\ifx \bparticle \undefined \def \bparticle#1{#1}\fi
\ifx \barticle \undefined \def \barticle#1{#1}\fi
\bibcommenthead
\ifx \bconfdate \undefined \def \bconfdate #1{#1}\fi
\ifx \botherref \undefined \def \botherref #1{#1}\fi
\ifx \url \undefined \def \url#1{\textsf{#1}}\fi
\ifx \bchapter \undefined \def \bchapter#1{#1}\fi
\ifx \bbook \undefined \def \bbook#1{#1}\fi
\ifx \bcomment \undefined \def \bcomment#1{#1}\fi
\ifx \oauthor \undefined \def \oauthor#1{#1}\fi
\ifx \citeauthoryear \undefined \def \citeauthoryear#1{#1}\fi
\ifx \endbibitem  \undefined \def \endbibitem {}\fi
\ifx \bconflocation  \undefined \def \bconflocation#1{#1}\fi
\ifx \arxivurl  \undefined \def \arxivurl#1{\textsf{#1}}\fi
\csname PreBibitemsHook\endcsname

\bibitem[\protect\citeauthoryear{{ENISA}}{2022}]{ENISA2022Threat}
\begin{botherref}
\oauthor{\bsnm{{ENISA}}}:
Threat {{Landscape Report}} 2022.
Technical report,
{European Union Agency for Cybersecurity}
(2022)
\end{botherref}
\endbibitem

\bibitem[\protect\citeauthoryear{Ahmad et~al.}{2021}]{ahmad2021network}
\begin{barticle}
\bauthor{\bsnm{Ahmad}, \binits{Z.}},
\bauthor{\bsnm{Shahid~Khan}, \binits{A.}},
\bauthor{\bsnm{Wai~Shiang}, \binits{C.}},
\bauthor{\bsnm{Abdullah}, \binits{J.}},
\bauthor{\bsnm{Ahmad}, \binits{F.}}:
\batitle{Network intrusion detection system: A systematic study of machine learning and deep learning approaches}.
\bjtitle{Transactions on Emerging Telecommunications Technologies}
\bvolume{32}(\bissue{1}),
\bfpage{4150}
(\byear{2021})
\end{barticle}
\endbibitem

\bibitem[\protect\citeauthoryear{Khraisat et~al.}{2019}]{khraisat2019survey}
\begin{barticle}
\bauthor{\bsnm{Khraisat}, \binits{A.}},
\bauthor{\bsnm{Gondal}, \binits{I.}},
\bauthor{\bsnm{Vamplew}, \binits{P.}},
\bauthor{\bsnm{Kamruzzaman}, \binits{J.}}:
\batitle{Survey of intrusion detection systems: techniques, datasets and challenges}.
\bjtitle{Cybersecurity}
\bvolume{2}(\bissue{1}),
\bfpage{1}--\blpage{22}
(\byear{2019})
\end{barticle}
\endbibitem

\bibitem[\protect\citeauthoryear{Aghaei et~al.}{2023}]{Aghaei2023SecureBERT}
\begin{bchapter}
\bauthor{\bsnm{Aghaei}, \binits{E.}},
\bauthor{\bsnm{Niu}, \binits{X.}},
\bauthor{\bsnm{Shadid}, \binits{W.}},
\bauthor{\bsnm{{Al-Shaer}}, \binits{E.}}:
\bctitle{{{SecureBERT}}: {{A Domain-Specific Language Model}} for~{{Cybersecurity}}}.
In: \beditor{\bsnm{Li}, \binits{F.}},
\beditor{\bsnm{Liang}, \binits{K.}},
\beditor{\bsnm{Lin}, \binits{Z.}},
\beditor{\bsnm{Katsikas}, \binits{S.K.}} (eds.)
\bbtitle{Security and {{Privacy}} in {{Communication Networks}}}.
\bsertitle{Lecture {{Notes}} of the {{Institute}} for {{Computer Sciences}}, {{Social Informatics}} and {{Telecommunications Engineering}}},
pp. \bfpage{39}--\blpage{56}.
\bpublisher{{Springer Nature Switzerland}}, \blocation{???}
(\byear{2023}).
\doiurl{10.1007/978-3-031-25538-0_3}
\end{bchapter}
\endbibitem

\bibitem[\protect\citeauthoryear{Akbar et~al.}{2022}]{Akbar2022Knowledge}
\begin{bchapter}
\bauthor{\bsnm{Akbar}, \binits{K.A.}},
\bauthor{\bsnm{Halim}, \binits{S.M.}},
\bauthor{\bsnm{Hu}, \binits{Y.}},
\bauthor{\bsnm{Singhal}, \binits{A.}},
\bauthor{\bsnm{Khan}, \binits{L.}},
\bauthor{\bsnm{Thuraisingham}, \binits{B.}}:
\bctitle{Knowledge {{Mining}} in~{{Cybersecurity}}: {{From Attack}} to~{{Defense}}}.
In: \beditor{\bsnm{Sural}, \binits{S.}},
\beditor{\bsnm{Lu}, \binits{H.}} (eds.)
\bbtitle{Data and {{Applications Security}} and {{Privacy XXXVI}}}.
\bsertitle{Lecture {{Notes}} in {{Computer Science}}},
pp. \bfpage{110}--\blpage{122}.
\bpublisher{{Springer International Publishing}},
\blocation{{Cham}}
(\byear{2022}).
\doiurl{10.1007/978-3-031-10684-2_7}
\end{bchapter}
\endbibitem

\bibitem[\protect\citeauthoryear{Buczak and Guven}{2016}]{Buczak2016survey}
\begin{barticle}
\bauthor{\bsnm{Buczak}, \binits{A.L.}},
\bauthor{\bsnm{Guven}, \binits{E.}}:
\batitle{A survey of data mining and machine learning methods for cyber security intrusion detection}.
\bjtitle{IEEE Communications Surveys Tutorials}
\bvolume{18}(\bissue{2}),
\bfpage{1153}--\blpage{1176}
(\byear{2016})
\doiurl{10.1109/COMST.2015.2494502}
\end{barticle}
\endbibitem

\bibitem[\protect\citeauthoryear{Lopez et~al.}{2017}]{lopez2017vector}
\begin{bchapter}
\bauthor{\bsnm{Lopez}, \binits{W.}},
\bauthor{\bsnm{Merlino}, \binits{J.}},
\bauthor{\bsnm{Rodriguez-Bocca}, \binits{P.}}:
\bctitle{Vector representation of internet domain names using a word embedding technique}.
In: \bbtitle{2017 XLIII Latin American Computer Conference (CLEI)},
pp. \bfpage{1}--\blpage{8}
(\byear{2017}).
\bcomment{IEEE}
\end{bchapter}
\endbibitem

\bibitem[\protect\citeauthoryear{Lopez et~al.}{2020}]{lopez2020learning}
\begin{barticle}
\bauthor{\bsnm{Lopez}, \binits{W.}},
\bauthor{\bsnm{Merlino}, \binits{J.}},
\bauthor{\bsnm{Rodriguez-Bocca}, \binits{P.}}:
\batitle{Learning semantic information from internet domain names using word embeddings}.
\bjtitle{Engineering Applications of Artificial Intelligence}
\bvolume{94},
\bfpage{103823}
(\byear{2020})
\end{barticle}
\endbibitem

\bibitem[\protect\citeauthoryear{Morbidoni et~al.}{2022}]{Morbidoni2022Leveraging}
\begin{bchapter}
\bauthor{\bsnm{Morbidoni}, \binits{C.}},
\bauthor{\bsnm{Spalazzi}, \binits{L.}},
\bauthor{\bsnm{Teti}, \binits{A.}},
\bauthor{\bsnm{Cucchiarelli}, \binits{A.}}:
\bctitle{Leveraging n-gram neural embeddings to improve deep learning {{DGA}} detection}.
In: \bbtitle{Proceedings of the 37th {{ACM}}/{{SIGAPP Symposium}} on {{Applied Computing}}},
pp. \bfpage{995}--\blpage{1004}.
\bpublisher{{ACM}},
\blocation{{Virtual}}
(\byear{2022}).
\doiurl{10.1145/3477314.3507269}
\end{bchapter}
\endbibitem

\bibitem[\protect\citeauthoryear{Roy et~al.}{2017}]{roy2017learning}
\begin{botherref}
\oauthor{\bsnm{Roy}, \binits{A.}},
\oauthor{\bsnm{Park}, \binits{Y.}},
\oauthor{\bsnm{Pan}, \binits{S.}}:
Learning domain-specific word embeddings from sparse cybersecurity texts.
arXiv preprint arXiv:1709.07470
(2017)
\end{botherref}
\endbibitem

\bibitem[\protect\citeauthoryear{{Cyber Edu}}{2018}]{CyberEdu2018What}
\begin{botherref}
\oauthor{\bsnm{{Cyber Edu}}}:
What Is the {{OSI Model}}?
https://www.forcepoint.com/cyber-edu/osi-model
(2018)
\end{botherref}
\endbibitem

\bibitem[\protect\citeauthoryear{Apruzzese et~al.}{2023}]{10190520}
\begin{bchapter}
\bauthor{\bsnm{Apruzzese}, \binits{G.}},
\bauthor{\bsnm{Laskov}, \binits{P.}},
\bauthor{\bsnm{Schneider}, \binits{J.}}:
\bctitle{Sok: Pragmatic assessment of machine learning for network intrusion detection}.
In: \bbtitle{2023 IEEE 8th European Symposium on Security and Privacy (EuroS\&P)},
pp. \bfpage{592}--\blpage{614}
(\byear{2023}).
\doiurl{10.1109/EuroSP57164.2023.00042}
\end{bchapter}
\endbibitem

\bibitem[\protect\citeauthoryear{Hastie et~al.}{2009}]{hastie2009support}
\begin{botherref}
\oauthor{\bsnm{Hastie}, \binits{T.}},
\oauthor{\bsnm{Tibshirani}, \binits{R.}},
\oauthor{\bsnm{Friedman}, \binits{J.}},
\oauthor{\bsnm{Hastie}, \binits{T.}},
\oauthor{\bsnm{Tibshirani}, \binits{R.}},
\oauthor{\bsnm{Friedman}, \binits{J.}}:
Support vector machines and flexible discriminants.
The elements of statistical learning: data mining, inference, and prediction,
417--458
(2009)
\end{botherref}
\endbibitem

\bibitem[\protect\citeauthoryear{Kim et~al.}{2024}]{kim2024flsec}
\begin{barticle}
\bauthor{\bsnm{Kim}, \binits{C.}},
\bauthor{\bsnm{So-In}, \binits{C.}},
\bauthor{\bsnm{Kongsorot}, \binits{Y.}},
\bauthor{\bsnm{Aimtongkham}, \binits{P.}}:
\batitle{Flsec-rpl: a fuzzy logic-based intrusion detection scheme for securing rpl-based iot networks against dio neighbor suppression attacks}.
\bjtitle{Cybersecurity}
\bvolume{7}(\bissue{1}),
\bfpage{27}
(\byear{2024})
\end{barticle}
\endbibitem

\bibitem[\protect\citeauthoryear{Di~Mauro et~al.}{2021}]{di2021supervised}
\begin{barticle}
\bauthor{\bsnm{Di~Mauro}, \binits{M.}},
\bauthor{\bsnm{Galatro}, \binits{G.}},
\bauthor{\bsnm{Fortino}, \binits{G.}},
\bauthor{\bsnm{Liotta}, \binits{A.}}:
\batitle{Supervised feature selection techniques in network intrusion detection: A critical review}.
\bjtitle{Engineering Applications of Artificial Intelligence}
\bvolume{101},
\bfpage{104216}
(\byear{2021})
\end{barticle}
\endbibitem

\bibitem[\protect\citeauthoryear{Fares et~al.}{2011}]{fares2011intrusion}
\begin{barticle}
\bauthor{\bsnm{Fares}, \binits{A.H.}},
\bauthor{\bsnm{Sharawy}, \binits{M.I.}},
\bauthor{\bsnm{Zayed}, \binits{H.H.}}:
\batitle{Intrusion detection: supervised machine learning}.
\bjtitle{Journal of Computing Science and Engineering}
\bvolume{5}(\bissue{4}),
\bfpage{305}--\blpage{313}
(\byear{2011})
\end{barticle}
\endbibitem

\bibitem[\protect\citeauthoryear{Wang and Stolfo}{2004}]{wang2004anomalous}
\begin{bchapter}
\bauthor{\bsnm{Wang}, \binits{K.}},
\bauthor{\bsnm{Stolfo}, \binits{S.J.}}:
\bctitle{Anomalous payload-based network intrusion detection}.
In: \bbtitle{Recent Advances in Intrusion Detection: 7th International Symposium, RAID 2004, Sophia Antipolis, France, September 15-17, 2004. Proceedings 7},
pp. \bfpage{203}--\blpage{222}
(\byear{2004}).
\bcomment{Springer}
\end{bchapter}
\endbibitem

\bibitem[\protect\citeauthoryear{Xie and Hu}{2013}]{xie2013evaluating}
\begin{bchapter}
\bauthor{\bsnm{Xie}, \binits{M.}},
\bauthor{\bsnm{Hu}, \binits{J.}}:
\bctitle{Evaluating host-based anomaly detection systems: A preliminary analysis of adfa-ld}.
In: \bbtitle{2013 6th International Congress on Image and Signal Processing (CISP)},
vol. \bseriesno{3},
pp. \bfpage{1711}--\blpage{1716}
(\byear{2013}).
\bcomment{IEEE}
\end{bchapter}
\endbibitem

\bibitem[\protect\citeauthoryear{Xu et~al.}{2024}]{xu2024procsage}
\begin{barticle}
\bauthor{\bsnm{Xu}, \binits{B.}},
\bauthor{\bsnm{Gong}, \binits{Y.}},
\bauthor{\bsnm{Geng}, \binits{X.}},
\bauthor{\bsnm{Li}, \binits{Y.}},
\bauthor{\bsnm{Dong}, \binits{C.}},
\bauthor{\bsnm{Liu}, \binits{S.}},
\bauthor{\bsnm{Liu}, \binits{Y.}},
\bauthor{\bsnm{Jiang}, \binits{B.}},
\bauthor{\bsnm{Lu}, \binits{Z.}}:
\batitle{Procsage: an efficient host threat detection method based on graph representation learning}.
\bjtitle{Cybersecurity}
\bvolume{7}(\bissue{1}),
\bfpage{51}
(\byear{2024})
\end{barticle}
\endbibitem

\bibitem[\protect\citeauthoryear{Choi et~al.}{2019}]{choi2019unsupervised}
\begin{barticle}
\bauthor{\bsnm{Choi}, \binits{H.}},
\bauthor{\bsnm{Kim}, \binits{M.}},
\bauthor{\bsnm{Lee}, \binits{G.}},
\bauthor{\bsnm{Kim}, \binits{W.}}:
\batitle{Unsupervised learning approach for network intrusion detection system using autoencoders}.
\bjtitle{The Journal of Supercomputing}
\bvolume{75},
\bfpage{5597}--\blpage{5621}
(\byear{2019})
\end{barticle}
\endbibitem

\bibitem[\protect\citeauthoryear{Binbusayyis and Vaiyapuri}{2021}]{binbusayyis2021unsupervised}
\begin{barticle}
\bauthor{\bsnm{Binbusayyis}, \binits{A.}},
\bauthor{\bsnm{Vaiyapuri}, \binits{T.}}:
\batitle{Unsupervised deep learning approach for network intrusion detection combining convolutional autoencoder and one-class svm}.
\bjtitle{Applied Intelligence}
\bvolume{51}(\bissue{10}),
\bfpage{7094}--\blpage{7108}
(\byear{2021})
\end{barticle}
\endbibitem

\bibitem[\protect\citeauthoryear{Vaiyapuri and Binbusayyis}{2020}]{vaiyapuri2020application}
\begin{barticle}
\bauthor{\bsnm{Vaiyapuri}, \binits{T.}},
\bauthor{\bsnm{Binbusayyis}, \binits{A.}}:
\batitle{Application of deep autoencoder as an one-class classifier for unsupervised network intrusion detection: a comparative evaluation}.
\bjtitle{PeerJ Computer Science}
\bvolume{6},
\bfpage{327}
(\byear{2020})
\end{barticle}
\endbibitem

\bibitem[\protect\citeauthoryear{Faber et~al.}{2022}]{faber2022active}
\begin{bchapter}
\bauthor{\bsnm{Faber}, \binits{K.}},
\bauthor{\bsnm{Corizzo}, \binits{R.}},
\bauthor{\bsnm{Sniezynski}, \binits{B.}},
\bauthor{\bsnm{Japkowicz}, \binits{N.}}:
\bctitle{Active lifelong anomaly detection with experience replay}.
In: \bbtitle{2022 IEEE 9th International Conference on Data Science and Advanced Analytics (DSAA)},
pp. \bfpage{1}--\blpage{10}
(\byear{2022}).
\bcomment{IEEE}
\end{bchapter}
\endbibitem

\bibitem[\protect\citeauthoryear{Zhauniarovich et~al.}{2018}]{zhauniarovich2018survey}
\begin{barticle}
\bauthor{\bsnm{Zhauniarovich}, \binits{Y.}},
\bauthor{\bsnm{Khalil}, \binits{I.}},
\bauthor{\bsnm{Yu}, \binits{T.}},
\bauthor{\bsnm{Dacier}, \binits{M.}}:
\batitle{A survey on malicious domains detection through dns data analysis}.
\bjtitle{ACM Computing Surveys (CSUR)}
\bvolume{51}(\bissue{4}),
\bfpage{1}--\blpage{36}
(\byear{2018})
\end{barticle}
\endbibitem

\bibitem[\protect\citeauthoryear{Choi et~al.}{2007}]{choi2007botnet}
\begin{bchapter}
\bauthor{\bsnm{Choi}, \binits{H.}},
\bauthor{\bsnm{Lee}, \binits{H.}},
\bauthor{\bsnm{Lee}, \binits{H.}},
\bauthor{\bsnm{Kim}, \binits{H.}}:
\bctitle{Botnet detection by monitoring group activities in dns traffic}.
In: \bbtitle{7th IEEE International Conference on Computer and Information Technology (CIT 2007)},
pp. \bfpage{715}--\blpage{720}
(\byear{2007}).
\bcomment{IEEE}
\end{bchapter}
\endbibitem

\bibitem[\protect\citeauthoryear{Grill et~al.}{2015}]{grill2015detecting}
\begin{bchapter}
\bauthor{\bsnm{Grill}, \binits{M.}},
\bauthor{\bsnm{Nikolaev}, \binits{I.}},
\bauthor{\bsnm{Valeros}, \binits{V.}},
\bauthor{\bsnm{Rehak}, \binits{M.}}:
\bctitle{Detecting dga malware using netflow}.
In: \bbtitle{2015 IFIP/IEEE International Symposium on Integrated Network Management (IM)},
pp. \bfpage{1304}--\blpage{1309}
(\byear{2015}).
\bcomment{IEEE}
\end{bchapter}
\endbibitem

\bibitem[\protect\citeauthoryear{Antonakakis et~al.}{2011}]{antonakakis2011detecting}
\begin{bchapter}
\bauthor{\bsnm{Antonakakis}, \binits{M.}},
\bauthor{\bsnm{Perdisci}, \binits{R.}},
\bauthor{\bsnm{Lee}, \binits{W.}},
\bauthor{\bsnm{Vasiloglou}, \binits{N.}},
\bauthor{\bsnm{Dagon}, \binits{D.}}:
\bctitle{Detecting malware domains at the upper dns hierarchy.}
In: \bbtitle{USENIX Security Symposium},
vol. \bseriesno{11},
pp. \bfpage{1}--\blpage{16}
(\byear{2011})
\end{bchapter}
\endbibitem

\bibitem[\protect\citeauthoryear{Wang et~al.}{2020}]{wang2020defending}
\begin{bchapter}
\bauthor{\bsnm{Wang}, \binits{D.}},
\bauthor{\bsnm{Wang}, \binits{P.}},
\bauthor{\bsnm{Zhou}, \binits{J.}},
\bauthor{\bsnm{Sun}, \binits{L.}},
\bauthor{\bsnm{Du}, \binits{B.}},
\bauthor{\bsnm{Fu}, \binits{Y.}}:
\bctitle{Defending water treatment networks: Exploiting spatio-temporal effects for cyber attack detection}.
In: \bbtitle{2020 IEEE International Conference on Data Mining (ICDM)},
pp. \bfpage{32}--\blpage{41}
(\byear{2020}).
\bcomment{IEEE}
\end{bchapter}
\endbibitem

\bibitem[\protect\citeauthoryear{Wu et~al.}{2022}]{wu2022graphbert}
\begin{bchapter}
\bauthor{\bsnm{Wu}, \binits{J.}},
\bauthor{\bsnm{Zhang}, \binits{C.}},
\bauthor{\bsnm{Liu}, \binits{Z.}},
\bauthor{\bsnm{Zhang}, \binits{E.}},
\bauthor{\bsnm{Wilson}, \binits{S.}},
\bauthor{\bsnm{Zhang}, \binits{C.}}:
\bctitle{Graphbert: Bridging graph and text for malicious behavior detection on social media}.
In: \bbtitle{2022 IEEE International Conference on Data Mining (ICDM)},
pp. \bfpage{548}--\blpage{557}
(\byear{2022}).
\bcomment{IEEE}
\end{bchapter}
\endbibitem

\bibitem[\protect\citeauthoryear{Vaswani et~al.}{2017}]{Vaswani2017Attention}
\begin{bchapter}
\bauthor{\bsnm{Vaswani}, \binits{A.}},
\bauthor{\bsnm{Shazeer}, \binits{N.}},
\bauthor{\bsnm{Parmar}, \binits{N.}},
\bauthor{\bsnm{Uszkoreit}, \binits{J.}},
\bauthor{\bsnm{Jones}, \binits{L.}},
\bauthor{\bsnm{Gomez}, \binits{A.N.}},
\bauthor{\bsnm{Kaiser}, \binits{{\L}.}},
\bauthor{\bsnm{Polosukhin}, \binits{I.}}:
\bctitle{Attention is all you need}.
In: \bbtitle{Proceedings of the 31st {{International Conference}} on {{Neural Information Processing Systems}}},
pp. \bfpage{6000}--\blpage{6010}
(\byear{2017})
\end{bchapter}
\endbibitem

\bibitem[\protect\citeauthoryear{Kipf and Welling}{2016}]{DBLP:journals/corr/KipfW16}
\begin{botherref}
\oauthor{\bsnm{Kipf}, \binits{T.N.}},
\oauthor{\bsnm{Welling}, \binits{M.}}:
Semi-supervised classification with graph convolutional networks.
CoRR
\textbf{abs/1609.02907}
(2016)
{\href{https://arxiv.org/abs/1609.02907}{{1609.02907}}}
\end{botherref}
\endbibitem

\bibitem[\protect\citeauthoryear{Devlin et~al.}{2019}]{devlin-etal-2019-bert}
\begin{bchapter}
\bauthor{\bsnm{Devlin}, \binits{J.}},
\bauthor{\bsnm{Chang}, \binits{M.-W.}},
\bauthor{\bsnm{Lee}, \binits{K.}},
\bauthor{\bsnm{Toutanova}, \binits{K.}}:
\bctitle{{BERT}: Pre-training of deep bidirectional transformers for language understanding}.
In: \bbtitle{Proceedings of the 2019 Conference of the North {A}merican Chapter of the Association for Computational Linguistics: Human Language Technologies, Vol. 1},
pp. \bfpage{4171}--\blpage{4186}.
\bpublisher{Association for Computational Linguistics},
\blocation{Minneapolis, Minnesota}
(\byear{2019}).
\doiurl{10.18653/v1/N19-1423} .
\burl{https://aclanthology.org/N19-1423}
\end{bchapter}
\endbibitem

\bibitem[\protect\citeauthoryear{Ester et~al.}{1996}]{ester1996density}
\begin{bchapter}
\bauthor{\bsnm{Ester}, \binits{M.}},
\bauthor{\bsnm{Kriegel}, \binits{H.-P.}},
\bauthor{\bsnm{Sander}, \binits{J.}},
\bauthor{\bsnm{Xu}, \binits{X.}}, \betal:
\bctitle{A density-based algorithm for discovering clusters in large spatial databases with noise}.
In: \bbtitle{Kdd},
vol. \bseriesno{96},
pp. \bfpage{226}--\blpage{231}
(\byear{1996})
\end{bchapter}
\endbibitem

\bibitem[\protect\citeauthoryear{Bahdanau et~al.}{2014}]{bahdanau2014neural}
\begin{botherref}
\oauthor{\bsnm{Bahdanau}, \binits{D.}},
\oauthor{\bsnm{Cho}, \binits{K.}},
\oauthor{\bsnm{Bengio}, \binits{Y.}}:
Neural machine translation by jointly learning to align and translate.
arXiv preprint arXiv:1409.0473
(2014)
\end{botherref}
\endbibitem

\bibitem[\protect\citeauthoryear{Howard and Ruder}{2018}]{howard-ruder-2018-universal}
\begin{bchapter}
\bauthor{\bsnm{Howard}, \binits{J.}},
\bauthor{\bsnm{Ruder}, \binits{S.}}:
\bctitle{Universal language model fine-tuning for text classification}.
In: \bbtitle{Proceedings of the 56th Annual Meeting of the Association for Computational Linguistics (Volume 1: Long Papers)},
pp. \bfpage{328}--\blpage{339}.
\bpublisher{Association for Computational Linguistics},
\blocation{Melbourne, Australia}
(\byear{2018}).
\doiurl{10.18653/v1/P18-1031} .
\burl{https://aclanthology.org/P18-1031}
\end{bchapter}
\endbibitem

\bibitem[\protect\citeauthoryear{Singh et~al.}{2019}]{9ync-vv09-19}
\begin{botherref}
\oauthor{\bsnm{Singh}, \binits{M.}},
\oauthor{\bsnm{Singh}, \binits{M.}},
\oauthor{\bsnm{Kaur}, \binits{S.}}:
TI-2016 DNS Dataset.
\doiurl{10.21227/9ync-vv09} .
\url{https://dx.doi.org/10.21227/9ync-vv09}
\end{botherref}
\endbibitem

\bibitem[\protect\citeauthoryear{Devlin}{2018}]{devlin2018bert}
\begin{botherref}
\oauthor{\bsnm{Devlin}, \binits{J.}}:
Bert: Pre-training of deep bidirectional transformers for language understanding.
arXiv preprint arXiv:1810.04805
(2018)
\end{botherref}
\endbibitem

\bibitem[\protect\citeauthoryear{Group}{2008}]{pcap_spec}
\begin{botherref}
\oauthor{\bsnm{Group}, \binits{T.T.}}:
Libpcap file format.
Technical report,
The Tcpdump Group
(2008).
\url{https://www.tcpdump.org/manpages/pcap-savefile.5.txt}
\end{botherref}
\endbibitem

\bibitem[\protect\citeauthoryear{Van~der Maaten and Hinton}{2008}]{van2008visualizing}
\begin{botherref}
\oauthor{\bsnm{Maaten}, \binits{L.}},
\oauthor{\bsnm{Hinton}, \binits{G.}}:
Visualizing data using t-sne.
Journal of machine learning research
\textbf{9}(11)
(2008)
\end{botherref}
\endbibitem

\end{thebibliography}

\end{document}